\documentstyle[12pt,aaspp,flushrt]{article}
\newcommand{\Ref}{\hangindent=20pt \hangafter=1 \noindent}
\newcommand{\StartRef}{\hyphenpenalty=10000 \raggedright}

\newcommand{\beq}{\begin{equation}}
\newcommand{\eeq}{\end{equation}}
\newcommand{\NarrowMargins}{
  \setlength{\oddsidemargin}{+0.3in}
  \setlength{\evensidemargin}{-0.0in}
  \setlength{\textwidth}{6.2in}
  \setlength{\topmargin}{-0.75in}
  \setlength{\textheight}{9.25in}   }
\catcode`\@=11 

\def\lsim{\mathrel{\mathpalette\@versim<}}
\def\gsim{\mathrel{\mathpalette\@versim>}}

\def\vp{v_{\parallel}}
\def\Ep{E_{\parallel}}
\def\om{\omega}
\def\omt{\tilde \omega}
\def\op{\Omega_p}
\def\rp{\rho_p}
\def\kp{k_{\perp}}
\def\lp{\lambda_p}

\def\@versim#1#2{\vcenter{\offinterlineskip
        \ialign{$\m@th#1\hfil##\hfil$\crcr#2\crcr\sim\crcr } }}
\catcode`\@=12 
\NarrowMargins
\input{psfig}
\begin{document}
\title{Particle Heating by Alfvenic Turbulence in Hot Accretion Flows} 
\author{Eliot Quataert\footnote{equataert@cfa.harvard.edu}}
\affil{Harvard-Smithsonian Center for Astrophysics,
60 Garden St., Cambridge, MA 02138}
\setcounter{footnote}{0}

\begin{abstract}
Recent work on Alfvenic turbulence by Goldreich \& Sridhar (1995; GS)
suggests that the energy cascades almost entirely perpendicular to the
local magnetic field.  As a result, the cyclotron resonance is
unimportant in dissipating the turbulent energy.  Motivated by the GS
cascade, we calculate the linear collisionless dissipation of Alfven
waves with frequencies much less than the proton cyclotron frequency,
but with perpendicular wavelengths of order the Larmor radius of
thermal protons.  In plasmas appropriate to hot accretion flows
(proton temperature $\gg$ electron temperature) the dissipated Alfven
wave energy primarily heats the protons.  For a plasma with $\beta
\lsim 5$, however, where $\beta$ is the ratio of the gas pressure to
the magnetic pressure, the MHD assumptions utilized in the GS analysis
break down before most of the energy in Alfven waves is dissipated; how
the cascade then proceeds is unclear.

Hot accretion flows, such as advection dominated accretion flows
(ADAFs), are expected to contain significant levels of MHD
turbulence. This work suggests that, for $\beta \gsim 5$, the Alfvenic
component of such turbulence primarily heats the protons.  Significant
proton heating is required for the viability of ADAF models.  We
contrast our results on particle heating in ADAFs with recent work by
Bisnovatyi-Kogan \& Lovelace (1997).\\

\noindent {\em Subject headings:} accretion -- hydromagnetics -- plasmas
-- turbulence
\end{abstract}

\section{Introduction}
In astrophysical accretion flows with significant angular momentum,
the accreting gas is believed to form either an optically thick,
geometrically thin, disk (Shakura \& Sunyaev 1973; see Frank, King, \&
Raine 1992 for a review) or an optically thin, geometrically thick,
quasi-spherical flow (Shapiro, Lightman, \& Eardley 1976; Ichimaru
1977; Rees et al. 1982; Narayan \& Yi 1994, 1995a, 1995b; Abramowicz
et al. 1995).  In thin accretion disks, the gas cools so efficiently
that all of the viscously generated energy is radiated locally.  The
resulting low temperature implies that the mean free path due to
Coulomb collisions is a small fraction of the size of the disk.
Consequently, a purely fluid description of the accreting gas is
reasonable.  By contrast, the high temperatures in hot accretion flows
entail that the (field-free) mean free path is often comparable to the
size of the accretion flow.  Collective plasma effects are thus likely
to be significant (Rees et al. 1982).

In hot accretion flows, the plasma is usually assumed to be two
temperature, with the ions significantly hotter than the electrons
(Shapiro, Lightman, \& Eardley 1976).  Such a temperature difference
is only possible if the ions and electrons are thermally decoupled.
For low accretion rates ($\dot m \lsim \alpha^2$, where $\dot m$ is
the accretion rate in Eddington units and $\alpha$ is the
Shakura-Sunyaev viscosity parameter) Coulomb collisions are too
inefficient to force a one temperature plasma (Rees et al. 1982). One
of the outstanding plasma physics problems relevant to accretion
theory is whether there are collective effects which transfer energy
from the ions to the electrons on a timescale short compared to the
inflow time of the gas, thus invalidating the two temperature
assumption (e.g., Phinney 1981).  The only such mechanism that we are
aware of in the astrophysical literature, due to Begelman \& Chiueh
(1988), is probably not efficient enough to eliminate the two
temperature nature of the flow (Narayan \& Yi 1995b).  In this paper
we thus assume that the two-temperature formalism is valid and that
the only thermal coupling between electrons and ions is due to Coulomb
collisions.

Recently, there has been substantial work (Narayan \& Yi 1994, 1995a,
1995b; Abramowicz et al. 1995; Chen et al. 1995; Nakamura et al. 1997)
on a class of hot optically thin accretion solutions first discovered
by Ichimaru (1977), advection--dominated accretion flows (ADAFs). The
defining characteristic of an ADAF is that it is underluminous for its
accretion rate (i.e., $L \ll {\dot M} c^2$).  This arises because
``standard'' ADAF models assume that, by virtue of their larger mass,
most of the viscously generated energy heats the ions.  Since only a
small fraction of this energy is transferred to the electrons via
Coulomb collisions, the total energy radiated (almost all by the
electrons) is much less than the total energy generated by viscosity
(Ichimaru 1977; Rees et al. 1982).  The remaining viscously generated
energy is stored as thermal energy of the ions and is advected onto
the central object.

The assumption that viscosity heats the ions is crucial for the
relevance of ADAF models.  This enables ADAF solutions to exist so
long as the electrons and ions are thermally decoupled ($\dot m \lsim
\alpha^2$).  If viscosity were to predominantly heat the electrons,
(optically thin) ADAF solutions would only exist when the electron
cooling time is longer than the inflow time of the gas, which occurs
provided $\dot m \lsim 10^{-4} \alpha^2$ for synchrotron cooling
(Mahadevan \& Quataert 1997).  Thus, in the context of ADAFs, the
issue of which particles receive the viscous energy acquires
particular importance.  In this paper, we present a preliminary
investigation of this question.  In \S6.5 we discuss some related work
by Bisnovatyi-Kogan \& Lovelace (1997).  We note that an investigation
similar to ours, but with somewhat different conclusions, was carried
out independently by Gruzinov (1997).

We assume that the energy generated by viscosity is initially
converted into large scale MHD waves.  For simplicity, we focus
primarily, but not exclusively, on Alfven waves.  The large scale
waves cascade to smaller wavelengths until they are dissipated.  The
relative heating of ions and electrons is determined by which particle
species is primarily responsible for the dissipation of the waves.  We
focus on linear collisionless dissipation mechanisms, as these should
be of principle importance in hot accretion flows (\S 6).

Collisionless dissipation of MHD waves and the back reaction of the
dissipated energy on the electron and proton distribution functions
has been considered extensively as a mechanism for accelerating
particles in solar flares (e.g., Melrose 1994; Miller \& Roberts 1995;
Miller, LaRosa \& Moore 1996).  These ideas have also recently been
applied to particle acceleration in accretion disk corona (Dermer,
Miller \& Li 1996; Li, Kusunose, \& Liang 1996; Li \& Miller 1997).
Following the seminal work of Kraichnan (1965), most of these
calculations assume that the turbulent cascade is isotropic, that is,
that the turbulent energy density at any scale depends only on the
magnitude of the wavevector, and is independent of its direction with
respect to the mean magnetic field.  Recent work on incompressible MHD
turbulence by Goldreich \& Sridhar (1995; 1997) and Sridhar \&
Goldreich (1994), collectively referred to hereafter as GS, suggests
that this assumption is inapplicable for Alfvenic turbulence. This is
discussed in more detail in the next section (\S 2).

We use the GS turbulent cascade to motivate the parameter regime (in
wavevector space) in which we investigate collisionless dissipation of
Alfven waves; we also briefly discuss the collisionless dissipation of
other MHD modes (\S 3).  In \S4 we qualitatively discuss the response
of the electron and proton distribution functions to the dissipation
of waves in a nearly perpendicular Alfvenic cascade (we do not,
however, solve the quasi-linear equations).  Throughout we concentrate
on plasmas for which the ion temperature is greater than (often much
greater than) the electron temperature and the gas pressure is
comparable to, or greater than, the magnetic pressure, a regime rarely
explored in calculations of the dissipation of MHD waves.  In \S5 we
apply our calculations to the GS cascade and address two difficult,
but important, questions: (1) to what extent is the linear analysis of
\S3 applicable to the strong Alfvenic cascade developed by GS and (2)
is the dissipation found in \S3 strong enough to dissipate the
turbulent energy before the MHD assumptions used in the GS analysis
are invalid.  In \S6 we apply our results to hot, two temperature,
accretion flows, in particular ADAFs, and in \S7 we summarize our
results.  We have attempted to make \S6, which contains our
astrophysical applications, comprehensible without a detailed
understanding of the plasma physics calculations in \S3-\S5.  Towards
this end, Appendix A contains definitions of a number of quantities
used repeatedly in this paper. 

\section{Alfvenic Turbulence}

Energy injected into a fluid/plasma on large spatial scales, if it is
unable to dissipate, builds up to nonlinear amplitudes and cascades to
smaller wavelength (larger wavevector) perturbations; this continues
until dissipation becomes important and the turbulent energy is
converted into thermal energy.  To investigate the range of
wavelengths where dissipation occurs (the ``dissipation range''
of the turbulence), two characteristics of the nonlinear cascade are
particularly important.  The first is the cascade time, i.e., the time
for nonlinear effects to transfer energy from a wavevector $\sim \bf
k$ to a wavevector $\sim 2 \bf k$.  This determines how rapid the
dissipation must be to halt the cascade.  The second is the path of
the cascade in wavevector space. Does it depend only on $|{\bf k}|
\equiv k$ or also on the direction?  If the dissipation is a function
of $\bf k$ (and not just $k$), as it is for MHD modes, this
distinction is crucial.

If there is no preferred direction in the fluid, the turbulence is
isotropic, i.e., just a function of $k$.  For MHD turbulence, however,
the local magnetic field picks out a direction and so isotropy is not
guaranteed.  In fact, numerical simulations have long shown that
incompressible MHD turbulence is anisotropic, with the energy
cascading primarily perpendicular to the mean magnetic field (e.g.,
Shebalin et al. 1983).  Incompressible MHD turbulence corresponds
roughly to cascading Alfven waves, since both the fast and slow MHD
modes are compressive.  Recent work on Alfvenic turbulence has
clarified the nature of this perpendicular cascade (GS; Montgomery \&
Matthaeus 1995; Ng \& Bhattacharjee 1996).

Linear Alfven waves satisfy the dispersion relation $\omega = v_A
k_z$, where $\omega$ is the mode frequency, $v_A$ is the Alfven speed
and $k_z$ is the component of the wavevector along the mean magnetic
field (taken to be along the z-direction).  GS argue that Alfvenic
turbulence naturally evolves into a ``critically balanced'' state in
which the cascade time at a scale $\bf k$ is comparable to the linear
wave period at that scale.  Furthermore, the parallel and
perpendicular sizes of a wave at any scale are correlated, with $k_z
\sim k^{2/3}_{\perp} L^{-1/3}$, where $L$ is the outer scale of the
turbulence (the scale on which energy is injected).  The fluctuating
magnetic field strength on any scale is given by $B_k \sim B_{out}
(\kp L)^{-1/3}$, where $B_{out}$ is the excitation amplitude on the
outer scale.\footnote{The GS analysis breaks down if the fluxes of
turbulent energy parallel and anti-parallel to the mean magnetic field
are not equal.  This precludes application of their theory to the
solar wind, but this assumption should be applicable in hot accretion
flows.}

In the next two sections we calculate properties of the linear
dissipation of nearly perpendicular Alfven waves ($\kp \gg k_z$).
This calculation is motivated by, but does not explicitly utilize, the
Alfvenic cascade of GS.  Because Alfven ``waves'' in the GS cascade
only live for $\sim$ a mode period before nonlinear effects transfer
their energy to smaller spatial scales, it is somewhat misleading to
speak of them as waves (the turbulence is strong rather than weak in
the turbulence theory sense).  We will generally treat this notion as
unproblematic, but in \S5, having discussed the results of the linear
analysis, we will return to the issue of its applicability to the GS
cascade.

\section{Linear Collisionless Dissipation of MHD Waves in Two Temperature 
Plasmas}
\subsection{Qualitative Considerations}

Waves in a magnetized plasma can in general have electric and magnetic
fields both perpendicular and parallel to the mean magnetic field.
These fields can strongly effect the motion of particles through
resonant interactions.  This occurs when the frequency of the wave, in
the frame moving with the particle along the field line, is an integer
multiple of the particle's cyclotron frequency, \beq \omega - k_z v_z
= n \Omega, \label{res} \eeq where $v_z$ is the particle's velocity
along the magnetic field and $\Omega$ is the relativistic
cyclotron frequency (e.g. Melrose 1980).  When this condition is
satisfied, the particle and wave are in phase and the wave can
efficiently accelerate the particle.  In a collision dominated plasma,
however, such phase coherence is impossible to maintain.

In the MHD limit, $\om \ll \op$ and $k \rp \ll 1$, where $\rp$ is the
Larmor radius of protons with the thermal speed and $\op$ is the
proton cyclotron frequency; in this limit the Alfven wave dispersion
relation is $\om = k_z v_A$ and the resonance condition becomes \beq
v_A - v_z = n v_A \Omega/\om. \label{resa} \eeq For $\om \ll \op$, $n
\ne 0$ resonances in equation (\ref{resa}) can only be satisfied by
particles with $v_z \sim n v_A \Omega/\om \gg v_A$. Since the thermal
speeds of particles are typically of order the Alfven speed, there are
a negligible number of such particles; $n \ne 0$ resonances are
consequently unimportant.

For $n = 0$, resonance occurs when the wave's phase speed along the
field line, $v_{\parallel} = \omega/k_z$, equals $v_z$.  Particles
with $v_z \lsim \vp$ are accelerated by the wave, while those with
$v_z \gsim \vp$ are decelerated.  Thus, the wave is damped provided
that the slope of the particle distribution function at $v_z = \vp$ is
negative -- as it is for a Maxwellian.  A necessary (but not
sufficient) condition for strong damping is that $\vp$ be comparable
to the thermal speed of the particles, so that there are a large
number of resonant particles.

The $n = 0$ resonance actually corresponds to two physically distinct
wave-particle interactions. In Landau damping (LD), particle
acceleration is due to the longitudinal electric field perturbation of
a wave (i.e., the usual electrostatic force, $E_z$).  In transit-time
damping (TTD), the magnetic analogue of LD, the interaction is between
the particle's effective magnetic moment ($\mu = m v^2_{\perp}/2 B$)
and the wave's longitudinal magnetic field perturbation, $B_z$ (Stix
1992).  TTD is thus analogous to Fermi acceleration (for discussions
of the relationship between the two, see Achterberg 1981; Miller
1991).

In the MHD limit, which corresponds to small wavevectors, the Alfven
wave has both $E_z = 0$ and $B_z = 0$ and so is undamped by linear
collisionless effects.  For larger wavevectors, the MHD approximations
are less applicable and kinetic theory corrections to $E_z$ and $B_z$
become important, leading to finite dissipation of the Alfven wave;
the $n \ne 0$ resonances {\em may} become important, if the wavevector
has a significant component parallel to the background magnetic field.
For the perpendicular cascade of Alfven waves due to GS, however, when
$\kp \rp \sim 1$, $\om \sim \op(\rp/ L)^{1/3} \ll \op$ and so $n \ne
0$ resonances can be satisfied only by particles with $v_z \gg v_A$,
of which there are a negligible number.  Since we expect significant
$E_z$ and/or $B_z$, and thus significant dissipation, when $\kp \rp
\sim 1$, this implies that, even in the dissipation range, only $n =
0$ resonances are important.

We now investigate in detail the collisionless dissipation of Alfven
waves, focusing on $\kp \rp \sim 1$ and $k_z \rp \ll 1$.  This problem
has been considered by a number of authors (e.g., Akhiezer et
al. 1975; Hasegawa \& Chen 1976; Stefant 1976), but not in the
parameter regime of interest to us.

\subsection{Detailed Calculations:  Methods}

Consider a collisionless hydrogen plasma which is homogeneous, fully
ionized, and threaded by a mean magnetic field, ${\bf B} = B_0
\bf{{\hat z}}$.  We assume that, in the unperturbed state, each
particle species in the plasma (electrons and protons) has an
isotropic, non-relativistic, thermal distribution function with no
bulk (average) velocities.\footnote{By isotropic we mean that the
temperature is the same perpendicular and parallel to ${\hat z}$.}
Small amplitude perturbations to the equilibrium state of the plasma
satisfy the following dispersion relation, which is obtained by
linearizing and Fourier transforming (in time and space) Maxwell's
equations (Stix 1992; Chapter 1) \beq {\bf k} \times ({\bf k} \times
{\bf E}) + {\om^2 \over c^2} {{ \epsilon } \cdot {\bf E}} = 0.
\label{disp} 
\eeq ${\bf E}$ is the electric field perturbation and the dielectric
tensor, ${\bf \epsilon}$, is given by $\epsilon_{ij} = \delta_{ij} +
\sum_s \chi^s_{ij}$, where the sum is over the susceptibility tensor
($\chi^s_{ij}$) of each particle species in the plasma; for our case,
$s$ takes on two values, e and p, for electrons and protons,
respectively.  The susceptibility tensor is calculated by combining
the linearized and Fourier transformed versions of Maxwell's equations
and the collisionless Boltzmann equation, and is given by, taking
${\bf k}$ in the x-z plane, (Stix 1992; Chapter 10)

$$ \chi^s_{xx} = {m_p v_{tp} \over m_s v_{ts}} 
{c^2 \over v^2_A} {\exp({-\lambda_s}) \over
\omt \lambda_s \eta} \sum_{n=-\infty}^{\infty} n^2 I_n Z_o(\xi_n)$$

$$\chi^s_{xy} = -\chi^s_{yx} = {m_p v_{tp} \over m_s v_{ts}} 
 {-i c^2 \over v^2_A} {\exp({-\lambda_s}) \over
\omt \eta} \sum_{n=-\infty}^{\infty} n (I_n -I^{'}_n) Z_o(\xi_n)$$

$$\chi^s_{yy} =  {m_p v_{tp} \over m_s v_{ts}}  
{c^2 \over v^2_A} {\exp({-\lambda_s}) \over
\omt \eta} \sum_{n=-\infty}^{\infty} \big[ {n^2 \over \lambda_s} I_n + 
2 \lambda_s (I_n -I^{'}_n)\big] Z_o(\xi_n)$$

$$\chi^s_{xz} = \chi^s_{zx} = {q_s \over |q_s|} {\sqrt 2 c^2 \over
v^2_A} {\sqrt \lambda_p \exp({-\lambda_s}) \over \omt \eta \lambda_s}
\sum_{n=-\infty}^{\infty} n I_n \big[ 1 + \xi_n Z_o(\xi_n) \big]$$

$$\chi^s_{yz} = - \chi^s_{zy} = {q_s \over |q_s|} {i \sqrt 2 c^2 \over v^2_A} {\sqrt \lambda_p
\exp({-\lambda_s}) \over \omt \eta} \sum_{n=-\infty}^{\infty} (I_n
-I^{'}_n)\big[1 + \xi_n Z_o(\xi_n)\big]$$

\beq
\chi^s_{zz} =  {m_p v^2_{tp} \over m_s v^2_{ts}} 
{2 c^2 \over v^2_A} {\exp({-\lambda_s}) \over \omt \eta^2} 
\sum_{n=-\infty}^{\infty} (\omt - n\Omega_s/\op) I_n 
\big[1 + \xi_n Z_o(\xi_n)\big], 
\label{susc}
\eeq where $v_{ts} = (2 k_B T_s/m_s)^{1/2}$, $q_s$, $m_s$, and $T_s$
are the thermal speed, charge, mass, and temperature of the particles,
respectively.  $v_A = B_0/(4 \pi \rho)^{1/2}$ is the Alfven speed and
$I_n$ is the modified Bessel function with argument $\lambda_s =
\kp^2 v^2_{ts}/2\Omega^2_s = 0.5 \kp^2 \rho^2_s$, where $\rho_s =
v_{ts}/\Omega_s$ is the Larmor radius of particles with the thermal
velocity and $\Omega_s = q_s B_0/m_s c$ is the non-relativistic
cyclotron frequency (taken to be a signed quantity).  Prime denotes
differentiation with respect to $\lambda_s$.  Note that we are
primarily interested in two temperature plasmas so that $T_p$ and
$T_e$ are, in general, not equal.

In deference to the problem of interest, we measure mode wavelengths
and frequencies in terms of the proton Larmor radius and cyclotron
frequency, using $\eta = k_z \rho_p$ and $\lambda_p = 0.5\kp^2\rp^2$
for the parallel and perpendicular components of the wavevector and
$\omt = \om/\Omega_p$ for the mode frequency.  We emphasize, however,
that equations (\ref{disp}) and (\ref{susc}) are valid regardless of
the magnitude of $\eta$, $\lambda_p$, or $\omt$, and can be used to
investigate the properties of any plasma waves (subject to the
validity of the assumptions stated at the beginning of this
subsection).

In equation (\ref{susc}), $Z_o$, the plasma dispersion function with
argument \beq \xi_n = {\om - n \Omega_s \over v_{ts} k_z}, \eeq is
given by, taking $k_z \ge 0$, (Stix 1992; Chapter 8) \beq Z_o(\xi) =
{1 \over \sqrt \pi} \int_{\Gamma} dz {\exp(-z^2) \over z - \xi},
\label{zo} \eeq where the contour, $\Gamma$, is such that the pole at
$z = \xi$ lies above the contour of integration in the complex $z$
plane.  Mathematically, collisionless dissipation arises from the
contribution of this pole to the susceptibility tensor, which is
$\propto \exp{(-\xi^2)}$ by the residue theorem.  Thus, a necessary
(but, again, not sufficient) condition for strong damping is $\xi_n
\lsim 1$, which is the thermal average of the single particle
resonance condition, equation (\ref{res}).

Solving the dispersion relation yields, for a given real $\bf k$, a
complex mode frequency, whose imaginary part, $\gamma$, represents the
growth or dissipation of the wave.  It does not, however, directly
reveal the relative energy absorbed by the protons and electrons as
the wave is damped.  Strictly speaking, this can only be obtained from
a nonlinear theory since energy considerations are necessarily second
order in the amplitude.  The Ohmic heating law, however, gives that
the rate of change of energy of particle species $s$ is $\propto {\bf
j_s \cdot E}$, where ${\bf j_s}$ is the current.  The complication is
that the Ohmic heating law includes both the increasing thermal energy
of the resonant particles (responsible for the damping) and the
decreasing oscillation energy of the non-resonant particles (e.g.,
Barnes 1968a).  When considering heating of the plasma by wave
dissipation, one wishes to calculate only the former, not the latter.
In the weak damping limit ($\gamma/\om \ll 1$), this can be obtained
from the Ohmic heating law by relating {$\bf j_s$} to {$\bf E$} using
the susceptibility ${\bf \chi}$ {\em evaluated at real frequencies}.
This yields (Barnes 1968b; Stix 1992) \beq P_s = {{\bf E^*} \cdot {\bf
\chi^a_s}|_{\rm Im(\om) = 0} \cdot {\bf E} \over 4 W}, \label{heat}
\eeq where $P_s$ is the energy absorbed in a mode period, per unit
wave energy, by particle species $s$ and $\chi^a_s = (\chi_s -
\chi^{\dagger}_s)/2i$ is the antihermitian part of the susceptibility
tensor.  Physically, $\chi^a_s$ is evaluated at real frequencies since
this entails that its only contribution is from the imaginary part of
$Z_o(\xi_n)$; in turn, for Im$(\om) = 0$, the only contribution to
Im$(Z_o)$ is from the poles at $z = \xi_n$.  As discussed below
equation (\ref{zo}), the contributions from the poles in $Z_o$
correspond to the thermal average of the single particle resonance
condition (eq. [\ref{res}]).  Setting Im$(\om) = 0$ in equation
(\ref{heat}) therefore isolates the contribution from the resonant
interactions, which is precisely what one wishes to do when
determining particle heating.  In equation (\ref{heat}), $W$, the wave
energy, is given by \beq W = {1 \over 16 \pi}\big[ |{\bf B}|^2 + {\bf
E^*} \cdot {\partial \over \partial \omega}(\omega \epsilon_h) \cdot
{\bf E} \big], \eeq where $\epsilon_h = (\epsilon +
\epsilon^{\dagger})/2$ is the hermitian part of the dielectric tensor
and $\bf B$ is the wave's magnetic field perturbation.  We note that,
for $\gamma/\omega \ll 1$, energy conservation implies that $P_p + P_e
= 2 \gamma T$, where $T \equiv 2 \pi /\rm Re(\om)$ is the mode period.
For the dissipation of Alfven waves of interest to us, this is well
satisfied even if $\gamma T \sim 1$; we therefore always use $P_p$ and
$P_e$ as estimates of the proton and electron heating rates,
respectively.

\subsection{Alfven Waves with large $\kp$}

Because equation (\ref{disp}) has an infinite number of roots (mostly
strongly damped waves with no fluid counterparts), some care must be
taken in its solution.  Our technique is to first solve the dispersion
relation in a simple limit (e.g., the MHD limit and plasma parameters
such that the wave is weakly damped); this ensures that we know which
mode we are investigating.  We then incrementally change the
wavevector and/or the plasma parameters and follow the properties of
the solution.  We emphasize that we have used the exact form of the
susceptibility tensor, with no approximations (save for terminating
the sum over Bessel functions at some appropriately large number).

In Figure 1 we show several properties of the Alfven wave as a
function of $T_p/T_e$ and the parameter $\lambda_p = 0.5 \kp^2 \rp^2$,
which measures the perpendicular wavelength of the wave; the results
correspond to a $\beta = 1$ plasma, where $\beta$, the ratio of the
gas pressure to the magnetic pressure, is given by $\beta = 8 \pi n
k_B (T_p +T_e)/B^2_0 = v_{tp}^2 (1 + T_e/T_p)/v^2_A$.  In the limit of
$k_z \rho_p \ll 1$ and $v_{ts},v_A \ll c$, which we take here, the
Alfven wave properties given in Figure 1 are independent of the exact
values of $k_z \rp$ and $v_{ts},v_A$, and depend only on $\lambda_p$,
$T_p/T_e$, and $\beta$.

Figure 1a shows the parallel phase speed of the Alfven wave in units
of the Alfven speed ($\vp/v_A \equiv {\rm Re}(\om)/k_z v_A$).  The two
curves correspond to $T_p = T_e$ and $T_p = 10^3 T_e$.  For $T_p \gsim
10 T_e$, $\vp$ is nearly identical to the $T_p = 10^3 T_e$ result
shown in the figure.  In the MHD limit ($\lambda_p \ll 1$), we have
$\vp \simeq v_A$, the usual Alfven wave dispersion relation, while in
the $ \lambda_p \gg 1$ limit $\vp \simeq v_A \sqrt \lambda_p \simeq
v_A \kp \rp$, i.e., the wave frequency depends strongly on the
perpendicular wave number.  This result is well-known from analytic
treatments in the $\beta \ll 1$ limit (e.g., Hasegawa \& Chen 1976).

The remaining panels in Figure 1 specify properties of the dissipation
of the Alfven wave by collisionless effects.  Figure 1b gives the
dissipation rate of the mode in units of the mode period, $\gamma T$.
Figure 1c gives the dimensionless proton, $P_p$, and electron, $P_e$,
heating rates, using equation (\ref{heat}).  For $T_p \gsim 10 T_e$,
the dimensionless proton heating rate, $P_p$, is nearly identical to
the $T_p = 10^3 T_e$ result shown in the figure.  We note that
$\gamma T$ (Fig. 1b) can be derived from $P_s$ (Fig. 1c) using $P_p +
P_e = 2 \gamma T$.  Finally, in Figure 1d we explicitly show the
relative heating of the protons and electrons, $P_p/P_e$.

As we are also interested in plasmas with $\beta \gsim 1$, Figure 2
shows $\gamma T$ (Fig. 2a) and $P_p/P_e$ (Fig. 2b) for several $\beta$
with $T_p/T_e$ = 100.  For $\beta \gsim 1$, the behavior of the
dissipation of the Alfven wave with varying $T_p/T_e$ is very similar
to that shown in Figure 1 (the $\beta = 1$ case): $\gamma$ is
relatively independent of $T_p/T_e$ for $\lambda_p \lsim 1$, but the
relative heating of protons and electrons ($P_p/P_e$) increases with
increasing $T_p/T_e$, as in Figure 1d.  This is explained analytically
below.

For our purposes, the relevant parameter regime in wavevector space is
$\lp \lsim 1$.  This is because, as is discussed in more detail in
\S5.2, the turbulent dynamics of Alfven waves due to GS is only valid in
this limit.  For $\lp \lsim 1$, the key qualitative results are that
the Alfven wave can be strongly damped by collisionless effects,
particularly if $\beta \gsim 1$.  Furthermore, provided that $T_p \gg
T_e$, the dissipated wave energy primarily heats the protons.  We note
that, for $\lp \gsim 1$, $\vp \simeq v_A \sqrt{\lp} \sim v_{tp}
\sqrt{\lp}$ and so with increasing $\lp$ there are progressively fewer
protons available to resonate with the wave.  This is why the proton
contribution to the damping, $P_p$, falls off exponentially for $\lp
\gsim 1$ (Fig. 1c).

The dissipation properties of the Alfven wave shown in Figures 1 and 2
can be understood in terms of primarily TTD by the protons; the
generally smaller electron contribution is due to both LD and TTD for
$\beta \simeq 1$, while TTD dominates for larger $\beta$.\footnote{For
$\beta \simeq 1$ and $T_p \simeq T_e$ the Landau and transit time
contributions are comparable for both electrons and protons.}  This
can be seen by looking at the relative contributions of TTD and LD in
the expression for the particle heating, $P_s$ (eq. [\ref{heat}]).
The TTD contribution is $\propto \chi^a_{yy} |E_y|^2$ (which is
$\propto \chi^a_{yy} |B_z|^2$ by Faraday's Law) while the LD term is
$\propto \chi^a_{zz} |E_z|^2$.  In Figure 3 we show, for $\lambda_p =
0.1$, the ratio of the TTD and LD contributions to the proton
(Fig. 3a) and electron (Fig. 3b) heating rates as a function of
$T_p/T_e$ for several $\beta$; below we discuss the
physical/analytical origin of these results.  We emphasize that, even
for $\lp \sim 1$, $\om \ll \op$ (since $\om \simeq k_z v_A$ and $k_z
\rp \ll 1$ by assumption); the magnetic moment of a particle is thus
an adiabatic invariant, as is required for TTD.

From equation (\ref{susc}), it is relatively straightforward to find
the dependence of $\chi_s^a$ on $\beta$ and $T_p/T_e$.\footnote{For
the regime of interest here, one need only keep the $n = 0$ terms and
the leading order $\lambda_p$ terms.} For a fixed $T_p/T_e$, this
yields $\chi^a_{yy}/\chi^a_{zz} \propto \beta$ for both protons and
electrons, while for a fixed $\beta$ we find that
$\chi^a_{yy,e}/\chi^a_{zz,e} \propto (T_e/T_p)^2$ and that
$\chi_{yy,p}^a/\chi^a_{zz,p}$ is independent of $T_p/T_e$.
Physically, increasing $\beta$ (at fixed $T_p/T_e$) increases the
proton and electron thermal speeds with respect to the Alfven speed,
thus increasing the magnetic moment of the thermal particles.  This
makes TTD more important, which is why $\chi^a_{yy}/\chi^a_{zz}
\propto \beta$.  Increasing $T_p/T_e$ (at fixed $\beta$), on the other
hand, effectively decreases the magnetic moment of the electrons,
which tends to decrease the contribution to electron heating from TTD.
This is why $\chi^a_{yy,e}/\chi^a_{zz,e} \propto (T_e/T_p)^2$.  The
parallel electric field of the Alfven wave, however, decreases with
increasing $T_p/T_e$.  In the $\beta \ll 1$ limit, $E_z \propto
T_e/T_p$ (at fixed $\beta$), while $E_y$ is independent of $T_e/T_p$
(e.g., Hasegawa \& Chen 1976; Melrose 1986, p. 178)\footnote{This is
because $E_y$ arises from keeping kinetic terms which are dropped in
the MHD limit while $E_z$ is due to both thermal and kinetic
corrections.}; we find that this is also roughly satisfied if
$\beta \simeq 1$.  These scalings imply that $\chi^a_{yy,p}
|E_y|^2/\chi^a_{zz,p} |E_z|^2 \sim \beta (T_p/T_e)^2$ and that
$\chi^a_{yy,e} |E_y|^2/\chi^a_{zz,e} |E_z|^2 \sim \beta$, which
reproduce the numerical calculations reasonably well.

These considerations show that, to a good degree of accuracy, one can
take $P_p/P_e \simeq \chi^a_{yy,p}/\chi^a_{yy,e}$, since both the
electron and proton heating is primarily due to TTD.  This greatly
simplifies evaluating $P_p/P_e$ analytically since a detailed
expression for the electric field vector is not needed.  Using
equation (\ref{susc}) and the MHD Alfven wave dispersion relation, we
find that \beq {P_p \over P_e} \simeq \left(m_p T_p \over m_e T_e
\right)^{1/2} \exp\left[-\left(1 + {T_e \over T_p}\right) \beta^{-1}
\right] \label{aheat}, \eeq where we have taken $v_{te} \gg v_A$ and
$\lp \lsim 1$.  The small deviations from equation (\ref{aheat}) in
Figures 1d and 2b (for $\lp \lsim 1$) are due to the contribution of LD
to the electron heating.  In the next section we give a more general
version of equation (\ref{aheat}) which is valid for any wave damped
primarily by TTD.

Finally, we note that the results given here for the dissipation of
nearly perpendicular Alfven waves in a $\beta \gsim 1$ plasma differ
from those obtained by Stefant (1976), who found that the Alfven wave
dissipation rate reaches a maximum ($\gamma T \simeq 0.1$) at some
particular value of $\beta$ and decreases for larger $\beta$. The
reason for this discrepancy is straightforward.  Stefant did not solve
the full dispersion relation (eq. [\ref{disp}]), but instead used a
simplified dispersion relation that neglected all contributions from
the $yy$ components of the susceptibility tensor.\footnote{His
dispersion relation also assumed $\lp \ll 1$.}  This amounts to
considering only LD of the Alfven wave.  This is valid only in the
$\beta \ll 1$ limit when the magnetic compression of the Alfven wave,
$B_z$, is negligible (e.g., Hasegawa \& Chen 1976, who use a
dispersion relation very similar to Stefant's, but explicitly state
that it is only valid for small $\beta$).  Our calculations agree with
Stefant's in the $\beta \ll 1$ limit; including the $yy$
susceptibilites, which are responsible for TTD, is, however, necessary
in the $\beta \sim 1$ limit.  In fact, in the limit of $\beta \gg 1$,
Foote and Kulsrud (1979) have shown analytically, by expanding the
susceptibility tensor to leading order in $\beta^{-1}$, that only TTD
contributes to the dissipation of the Alfven wave.

\subsection{General Relations for Particle Heating by TTD and LD}

Equation (\ref{aheat}) for the relative heating of protons and
electrons in the dissipation of a nearly perpendicular Alfven wave is
more general than it might appear.  From the susceptibility tensor
(eq. [\ref{susc}]) and the particle heating rate (eq. [\ref{heat}])
one can show that, for any wave which is damped solely by TTD, the
relative heating of protons and electrons in the $\lp \lsim 1$ limit
is \beq \left({P_p \over P_e} \right)_{TTD} \simeq \left(m_p T_p \over
m_e T_e \right)^{1/2} \exp\left[-\left({\vp \over v_{tp}}\right)^2 +
\left({\vp \over v_{te}}\right)^2 \right] \simeq \left(m_p T_p \over m_e
T_e \right)^{1/2}
\label{TTD}.  \eeq A similar analysis for LD shows that, for any wave
which is damped solely by LD, the relative heating of protons and
electrons in the $\lp \lsim 1$ limit is \beq \left({P_p \over P_e}
\right)_{LD} \simeq \left(m_p T_e^3 \over m_e T_p^3 \right)^{1/2}
\exp\left[-\left({\vp \over v_{tp}}\right)^2 + \left({\vp \over
v_{te}}\right)^2 \right] \simeq \left(m_p T_e^3 \over m_e T_p^3
\right)^{1/2}
\label{LD}. \eeq The latter equalities in equations (\ref{TTD}) and
(\ref{LD}) correspond to either subthermal waves ($\vp \lsim
v_{te},v_{tp}$) or equal electron and proton thermal speeds ($v_{te}
\simeq v_{tp}$).  The physical origin of these relations is as
follows.  The LD expression can be written as $(P_p/P_e)_{LD} \simeq
(m_e/m_p)(v^3_{te}/v^3_{tp})$.  The first term ($m_e/m_p$) is the
relative acceleration of protons and electrons for a given force.  The
second term ($v^3_{te}/v^3_{tp}$) is the relative number of particles
available to resonate with the wave (i.e., the relative slopes of the
proton and electron distribution function at the wave's phase speed,
taking $\vp \lsim v_{ve},v_{tp}$).  Equation (\ref{TTD}) for TTD can
be written as $(P_p/P_e)_{TTD} \simeq
(m_e/m_p)(v^3_{te}/v^3_{tp})(\mu^2_p/\mu^2_e)$, where $\mu_s \propto
T_s$ is the magnetic moment of particles with the thermal velocity.
The first two terms in the TTD expression are identical to the LD
expression, and have the same physical interpretation.  The last term
reflects the fact that in TTD, the wave-particle interaction is a
function of the particle's magnetic moment.  The larger $\mu_s$, the
stronger the coupling between the wave and the particle.  In LD the
corresponding term is $q_p^2/q_e^2 = 1$ since the wave-particle
coupling is the electrostatic force.

Equation (\ref{TTD}) shows that, quite generally, TTD is a natural
mechanism for preferentially heating protons in plasmas with $T_p \gg
T_e$.  This is because in such plasmas the protons have the larger
magnetic moment and so couple better to a wave's magnetic field
perturbation.  LD, on the other hand, leads to preferential electron
heating in plasmas with $T_p \gg T_e$.

\subsection{Collisionless dissipation of the fast and slow MHD modes}

A given excitation at the outer scale will, in general, contain both
noncompressive (Alfvenic) and compressive (fast and slow mode)
components.  For completeness, we therefore briefly consider the
collisionless dissipation of the fast and slow MHD modes in plasmas
with $T_p \gg T_e$.  Since these modes have either $B_z \ne 0$ or $E_z
\ne 0$ in the MHD limit (and as there is no detailed theory of fast or
slow mode turbulence to indicate if the cascade is parallel,
perpendicular, or isotropic) we consider the dissipation only in the
MHD limit.  For $T_p \simeq T_e$ this problem has been considered in
detail by Barnes (1966; 1967; 1968a; 1968b).  Figure 4 shows the
dimensionless dissipation rate, $\gamma T$, of the fast (Fig. 4a) and
slow (Fig. 4b) MHD modes as a function of $\theta$, the angle between
the wavevector and the background magnetic field, for several
$T_p/T_e$ for a $\beta = 1$ plasma (we take $k \rp \ll 1$ and
$v_A,v_{ts} \ll c$).  The $T_p = T_e$ results are identical to those
of Barnes (1966).

It is well known that the fast mode is damped primarily by TTD (Barnes
1966; Miller 1991).  This is because, in the MHD limit, the electric
field vector of the wave is along ${\bf \hat y}$, which leads to a
strong compressional magnetic field perturbation, $B_z$.  The relative
heating of the protons and electrons is thus given by equation
(\ref{TTD}) of the previous section: \beq \left({P_p \over P_e}
\right)_{fast} \simeq \left(m_p T_p \over m_e T_e \right)^{1/2}
\exp\left[-\left(1-{T_p m_e \over T_e m_p}\right)\left(1 + {T_e \over
T_p}\right)\sec^2(\theta) \right]
\label{fast}, \eeq where we have taken $\beta \simeq 1$ and have used the 
MHD dispersion relation $\omega \simeq k v_A$ (which is also
reproduced by our kinetic theory calculations), so that $\vp \simeq
\sec(\theta) v_A$.  In a one temperature plasma, electrons are
preferentially heated by fast modes with $\theta \gsim
60^{\circ}$. This is because, at these angles, $v_{te} \gg \vp \gg
v_{tp}$ and there are no resonant protons, but plenty of resonant
electrons (see also eq. [\ref{fast}]).  In a $T_p \gg T_e$ plasma,
however, there is no propagation angle for which the fast mode is
strongly damped and the electrons are preferentially heated.  This is
because $v_{te} \sim v_{tp}$ and so the above condition on the wave's
parallel phase speed cannot be obtained.  This also follows directly
from equation (\ref{fast}) by setting $v_{tp} \sim v_{te}$, in which
case $(P_p/P_e)_{fast} \simeq (m_p T_p/ m_e T_e)^{1/2}$, independent
of $\theta$.  In the MHD regime, fast mode turbulence in plasmas with
$T_p \gg T_e$ should therefore lead to primarily proton heating. There
is, however, a large range of propagation angles ($\theta \gsim
60^{\circ}-70^{\circ}$; see Fig. 4a) for which the fast mode is
essentially undamped (as opposed to $\theta \gsim 88^{\circ}$ for a
one temperature plasma).  These modes would likely cascade out of the
MHD regime.

The slow MHD mode is essentially a sound wave modified by the presence
of a magnetic field; it thus has a large $E_z$ and is Landau damped.
For $T_p \gg T_e$ this leads to preferential electron heating (\S3.4).
As Figure 4b indicates, the slow mode is very strongly damped by
collisionless effects; this is particularly true in a $T_p \gg T_e$
plasma, since there are more electrons available to resonate with the
wave.  This likely precludes slow mode turbulence cascading out of the
MHD regime (again, particularly in a $T_p \gg T_e$ plasma).

\section{The Effect of Wave Dissipation in a Perpendicular Alfvenic Cascade
on the Electron and Proton Distribution Functions}

Particle acceleration by Alfvenic turbulence has often been discussed
utilizing either parallel or isotropic turbulent cascades (e.g., for
solar flares, Miller \& Roberts 1995; Smith \& Miller 1995; for
accreting black holes, Dermer et. al. 1996; Li et. al. 1996).  In the
case of an isotropic cascade, these works also neglect the $n = 0$
wave particle interactions (which are the focus of this paper).  These
assumptions allow a dramatic simplification of the quasi-linear
diffusion equations, which describe the response of the particles to
the dissipated wave energy as a diffusion in velocity space (e.g.,
Melrose 1980).  In contrast to the perpendicular cascade of Alfven
waves considered here, dissipation occurs when $\om \sim \op$.  This
corresponds to the $|n| = 1$ resonance in the single particle resonance
equation (eq. [\ref{res}]).  In this case the turbulent energy is
transferred entirely to the protons, regardless of $T_p/T_e$ (since
the damping is due to a resonance between the proton's cyclotron
motion and the wave's perpendicular electric field).

Dissipation of Alfvenic turbulence by the cyclotron resonance is
attractive because it naturally leads to the formation of strong
non-thermal features in the proton distribution function (and
acceleration of protons to relativistic energies), which are suggested
by observations of both solar flares and accreting black holes.  To
see qualitatively how this occurs, note that an isotropic Alfvenic
cascade has waves with frequencies from the outer scale frequency up
to $\sim \op$.  From the resonance condition, equation (\ref{res}), we
see that waves with $\om \simeq \op$ accelerate protons with $v_z \sim
v_A \sim v_{tp}$ (we take $\beta \sim 1$ in this section).  Waves with
$\om \lsim \op$ can only accelerate particles if there is a big
Doppler shift in the wave frequency, i.e., particles with $v_z \sim
v_{tp} \op/\om$.  Thus a spectrum of waves with $\om/\op$ ranging from
$\ll 1$ up to $\sim 1$ can naturally accelerate particles from thermal
to relativistic energies.  The above references contain detailed
calculations of the evolution of the proton distribution function by
this process.

The work of GS suggests, however, that isotropic Alfvenic cascades are
unlikely to be obtained.  As we noted in \S3.1, when dissipation in a
nearly perpendicular Alfvenic cascade occurs, $\om \ll \op$. This
implies that $n \ne 0$ resonances (such as the cyclotron resonance)
are unimportant since there are a negligible number of particles with
$v_z \sim v_{tp} \op/\om \gg v_{tp}$.  Since only $n = 0$ resonances
are relevant, the resonance condition simplifies to $\om = k_z v_z$
or, using the MHD Alfven wave dispersion relation, $v_z = v_A \sim
v_{tp}$.  We emphasize that this condition, which is independent of
the wave frequency and wavevector, holds for {\em all waves} in a
nearly perpendicular Alfvenic cascade.  The effect of the turbulent
energy on the proton distribution function is thus qualitatively as
follows.  All waves dissipate their energy to particles with $v_z =
v_A \sim v_{tp}$.  This increases the {\em parallel} energy of
particles near the peak of the thermal distribution, but no
suprathermal feature is formed; there are simply no waves which can
accelerate suprathermal particles.  Only the parallel energy of the
particles increases since, for $\om \ll \op$, a particle's magnetic
moment is an adiabatic invariant; its perpendicular energy therefore
cannot change.  The resulting parallel proton distribution function is
reasonably well approximated as thermal, since it is roughly monotonic
and has a well-defined mean energy (e.g., Begelman \& Chiueh 1988).
For $T_p \gg T_e$, the same holds for electrons since then $v_{te}
\sim v_A$ and the particles which are heated also reside near the peak
of the thermal distribution function.\footnote{For $T_p \sim T_e$, the
electron heating may lead to a ``bump'' in the distribution function
at $v_z \simeq v_A \ll v_{te}$ (in addition to the thermal peak near
$v_{te}$).  The same holds for protons if $\beta$ is much different
from 1 since then $v_{tp} \ne v_A$.  In these cases, there is still no
acceleration of particles to relativistic energies, although a
potentially significant non-thermal (non-monotonic) feature in the
distribution function may develop (depending on the ratio of the
turbulent energy dissipated to the thermal energy near $v_z \simeq
v_A$).}

The fact that TTD only heats the parallel energy of the particles is
significant since the efficiency of TTD depends on the particle's
magnetic moment, which is proportional to the particle's perpendicular
energy.  If $E_{\perp}$ is unchanged as the plasma is heated the
efficiency of TTD would be substantially reduced.  In \S6.4 we discuss
mechanisms which lead to isotropy in the distribution function in hot
accretion flows.

\section{Applications to the GS cascade}

\subsection{Validity of the linear theory}

For the linear analysis of the previous sections to be applicable, a
particle's motion must be reasonably well approximated by the guiding
center approximation (free streaming along $\bf B_o$ and cyclotron
motion perpendicular to $\bf B_o$).  The presence of a turbulent
cascade clearly perturbs a particle's motion, so it is worthwhile
examining to what extent the linear analysis is valid for the strong
turbulent cascade of GS.  First we consider motion parallel to the
mean field, which is particularly important since we are interested in
the $n = 0$ resonances, which occur between the free streaming motion
of the particle and the wave's parallel fields.

For large wavevectors the Alfven wave has finite perturbed electric
and magnetic fields along the mean magnetic field ($E_z$ and $B_z$),
which are responsible for the damping discussed in the previous
sections.  If the parallel fields are too large, however, they trap
the particles in the potential fluctuations of the wave, violating the
assumption of free streaming.  This occurs within the cascade time
provided that $\om \tau_{e,b} \lsim 1$, where $\tau_{e} \sim (m/qk_z
E_z)^{1/2}$ is the characteristic oscillation period in the potential
well of the parallel electric field (Stix 1992) and $\tau_b \sim
(m/\mu k_z^2 B_z)^{1/2}$ is its analogue for the parallel magnetic
field.  From the numerical calculations, $E_z \sim E_x \lp^{1/2} \eta
T_e/T_p$ and $E_y \sim E_x \eta$ (taking $\beta \sim 1$; see also
Melrose 1986, p. 178); using Faraday's Law and the GS scaling for $B_y
= B_k$ from \S2, \beq E_z \sim B_{out} {v_A \over c} {T_e \over T_p}
\lp^{2/3} \left({\rp \over L}\right)^{2/3}
\label{ez} \eeq and \beq B_z \sim B_{out} \lp^{1/3} \left({\rp \over
L}\right)^{1/3}. \label{bz} \eeq Even though the turbulent energy
density ($B^2_k$) decreases with increasing $\bf k$ (\S2), the
parallel field strengths increase with increasing $\bf k$ because the
kinetic theory corrections win out over the decreasing wave energy.
Using equations (\ref{ez}) and (\ref{bz}), it is straightforward to
check that, for the protons, $\om \tau_{e} \sim \lp^{-1/6}
(L/\rp)^{1/6} (T_p/T_e)^{1/2} \gg 1$ and $\om \tau_b \sim \lp^{-1/6}
(L/\rp)^{1/6} \gg 1$ (for the applications discussed in \S6, $L/\rp
\gsim 10^8$).  Thus, the turbulence does not appear to contain
sufficiently large parallel fields to trap the particles.

We now consider motion perpendicular to the mean field, which is more
complicated.  For a given wave in the GS cascade the ${\bf E \times
B}$ drift velocity is ${\bf v_E } = c {\bf E \times B}/B^2 \sim v_A
(\kp L)^{-1/3}( -{\bf \hat y} + E_y/E_x {\bf \hat x})$.  Using the
above scaling for $E_y/E_x$ implies \beq {\bf v_E} \sim v_A \left[
-\lp^{-1/6} \left({\rp \over L} \right)^{1/3} {\bf \hat y} + \lp^{1/6}
\left({\rp \over L}\right)^{2/3} {\bf \hat x} \right] . \label{ve}
\eeq This leads to a Doppler shift in the frequency of the wave seen
by the particle of ${\bf k \cdot v_E} = \kp v_{E,x} \sim \om \lp^{1/3}
(\rp/L)^{1/3}$, since $\bf k$ is in the x-z plane (\S3.2).  In
addition, the finite lifetime of Alfven waves in the GS cascade
introduces shifts in the mode frequency which are $\sim \omega$ (since
the cascade time is of order the linear mode period).

Since the Doppler shifts due to the GS cascade are $\sim \om \ll \op$,
the magnetic moment of the particles is still an adiabatic invariant.
This is important for the applicability of TTD.  The single particle
resonance condition (eq. [\ref{res}]), however, shows up as a delta
function in the linear theory. It therefore cannot be rigorously
applicable to the GS cascade since there are additional frequency
shifts $\sim \om$ not accounted for in the linear theory.  This leads
to a broadening of the resonance condition, but only by a factor of
order unity.  The usual result of resonance broadening is that it
makes more particles available to resonate with the wave.  For our
problem, this has little effect since in linear theory there are
already a significant number of resonant particles -- the parallel
phase speed of the Alfven wave is comparable to the proton and
electron thermal velocities.  Furthermore, while the turbulence is
strong in a turbulence theory sense, the wave amplitudes are not so
large as to broaden the resonance condition by many harmonics of the
wave frequency, which would significantly reduce the particle heating
(Begelman \& Chiueh 1988).\footnote{This happens when the Doppler
shifts are much greater than the mode frequency.}  These
considerations suggest to us that the linear analysis should be a good
first approximation for the GS cascade.  This does not, of course,
preclude that other dissipation mechanisms neglected in the linear
analysis could be important.

\subsection{Resolution of the GS Cascade}

In the GS cascade, the cascade time is of order the linear Alfven
period; conversion of a significant fraction of the turbulent energy
to thermal energy via dissipation thus requires a damping time of
order the mode period, i.e., $\gamma T \sim 1$.  Furthermore, to
consistently apply the turbulent dynamics of GS, this dissipation must
occur when the MHD approximations made in their calculation are
reasonably applicable.  This requires $k \rp \lsim 1$ and $\om \lsim
\op$, which simplify to $\kp \rp \lsim 1$ for the nearly perpendicular
cascade of GS.

From the calculations described in \S3, it follows that (for $\beta
\gsim 1$ plasmas) Alfven waves with $\kp \rp \lsim 1$ and $\om \ll
\op$ have damping rates satisfying $\gamma T \gsim 1$ only if $\beta
\gsim 5$.  In large $\beta$ plasmas the proton magnetic moment couples
better to the wave's magnetic field perturbation, leading to more
efficient TTD.  If $\beta \gsim 5$, most ($\gsim 50 \%$) of the
turbulent energy in Alfven waves is dissipated in the GS cascade. The
relative heating of protons and electrons by this dissipated energy is
given approximately by equation (\ref{aheat}).

What happens, however, for the astrophysically important case of a
$\beta \sim 1$ plasma?  Proton heating still dominates over electron
heating, but Figure 1 gives $\gamma T \simeq 0.1$ for an Alfven wave
with $\kp \rp \sim 1$, so that only a small fraction of the Alfvenic
energy is dissipated.  TTD is not strong enough to damp the waves
before they cascade out of the MHD regime.  We can envision three
possibilities for how the energy is ultimately dissipated, but are
unable to ascertain which is realized.

{\noindent 1.  Other dissipation mechanisms are important and
dissipate the wave energy before $\kp \gsim \rp^{-1}$.}

{\noindent 2.  The cascade continues past $\kp \sim \rp^{-1}$, but how
it does so (i.e., along what track in $\bf k$ space) and in which
modes the energy resides is unknown; how the turbulent energy is
ultimately dissipated is thus unknown.  The structure of the GS
cascade is crucially dependent on the polarization (i.e., the velocity
eigenfunction) and dispersion relation of the Alfven wave.  Both of
these properties are entirely different for $\kp \rp \gsim 1$ (the
kinetic limit) than they are in the GS regime ($\kp \rp \lsim 1$).
For example, the wave frequency is $\propto \kp$ in the kinetic limit
while it is independent of $\kp$ in the GS regime.  In the GS regime,
the electrons and protons have the same velocity structure (the {$\bf
E \times B$} drift is independent of $q$ and $m$), while in the
kinetic limit, the electrons move substantially faster than the
protons;\footnote{This is because the wave's electric field averages
out over the proton Larmor orbit, but not the electron's; the protons
thus see an effectively smaller electric field and move more slowly as
a result.} a single fluid analysis is thus no longer applicable.  It
is therefore unclear whether the turbulent energy will stay in the
``Alfven'' wave in the kinetic limit.  Furthermore, GS argue that, in
the $\kp \rp \lsim 1$ limit, the Alfven wave in the nearly
perpendicular cascade is poorly coupled to the slow and fast modes;
this conclusion also rests on the polarization and dispersion
relations of the modes and is thus inapplicable in the kinetic limit.
For $\kp \rp \gsim 1$, the Alfven wave therefore may (or may not)
efficiently couple to fast or slow (or other) modes.  Understanding
this possibility is clearly a rather complicated problem in plasma
turbulence. }

{\noindent 3. As a result of poor coupling to other waves, the energy
is dissipated at $\kp \rp \sim 1$.  As in the previous possibility,
the energy ``wants'' to cascade to higher $\kp$, but being poorly
coupled to any other modes, it can't; the cascade time and mode energy
thus increase (maintaining a constant energy flux) until the waves are
dissipated.}

\section{Applications to Hot, Two Temperature, Accretion Flows}

\subsection{General Considerations}

Angular momentum transport in thin accretion disks is now believed to
arise from a magneto-rotational MHD instability discussed extensively
by Balbus and Hawley (1991, hereafter BH; for a recent review, see
Balbus \& Hawley 1997). This instability has been considered primarily
in the MHD limit, and its applicability to nearly collisionless
systems is perhaps unclear.  In Appendix B we argue (but do not prove)
that the instability should proceed in collisionless systems, provided
that the particle distribution functions are close to thermal.  The BH
instability, when it reaches nonlinear amplitudes, is large scale ($k
\sim H^{-1} \sim R^{-1}$, where $k$ is the wavevector of the
instability, $H$ is the disk scale height, $R$ is the local radius in
the accretion flow, and the latter equality is for quasi-spherical
accretion flows) and so naturally couples to long wavelength waves,
generating MHD turbulence; this is seen in numerous numerical
simulations (e.g., Stone et al. 1996).  This is the basic reason for
supposing that most of the gravitational potential energy released by
viscosity resides in MHD turbulence.\footnote{More generally, any
large scale instability in the plasma will generate significant levels
of MHD turbulence.}  In this paper we have made the further
simplification of focusing primarily on Alfvenic turbulence; this is
both because Alfvenic turbulence is (comparably) well understood and
because some such restriction is a necessary first step in attempting
to understand the problems considered in this paper.  MHD turbulence
generated by the BH instability may in fact be predominantly Alfvenic
since the instability is, to linear order, noncompressive.  This
suggests that the excitation of compressive MHD turbulence (such as
fast and slow modes) may be less important since it is a higher order
nonlinear effect than the excitation of noncompressive (Alfven)
modes.\footnote{The Alfven wave in a compressible medium is
incompressive to linear order, but compressive when nonlinear effects
are included (Holwegg 1971).}

Advection dominated accretion flows (ADAFs) are the only known
thermally and viscously stable, dynamically consistent models of hot,
two temperature, accretion flows (Kato et al. 1996).  For this reason,
we frame the discussion in this section in terms of ADAFs.  Much of
what we say, however, will apply to any hot, two temperature, accretion
flow.  For the present purposes, the relevant properties of ADAFs are
well described by the self similar solution of Narayan \& Yi (1994;
1995b), which yields
$${v \over c } \simeq  0.37 \, {\alpha}  \, r^{-1/2}, $$
$$T_p \simeq 2 \times 10^{12} {\beta \over \beta + 1}r^{-1} \ \ \mbox{K},$$
$$ n \simeq 6.3 \times 10^{19} \, 
		{\alpha^{-1} }  \, m^{-1} \dot{m} \, 
		r^{-3/2} \ \ \mbox{cm$^{-3}$},  $$ 
$$
B \simeq 10^9 \, \alpha^{-1/2} 
		\left( {\beta + 1} \right)^{-1/2} \, 
	m^{-1/2}\dot{m}^{1/2} \, r^{-5/4} \ \ \mbox{Gauss}, $$
$$
\op \simeq 10^{13} \alpha^{-1/2} \left( {\beta + 1} \right)^{-1/2} \,
m^{-1/2}\dot{m}^{1/2} \, r^{-5/4} \ \ \mbox{rad s$^{-1}$}, $$ \beq
{\rp \over R} \simeq 6 \times 10^{-9} \beta^{1/2} \alpha^{1/2}
m^{-1/2} {\dot m}^{-1/2} r^{-1/4}, \label{adaf} \eeq where $v/c $ is
the radial velocity in units of the velocity of light, $n$ is the
number density of electrons/protons, $B$ is the magnetic field
strength, determined by assuming a constant $\beta$ in the accretion
flow\footnote{The ``$\beta$'' used in papers on ADAF models,
$\beta_{\rm adv}$, is taken to be the ratio of the gas pressure to the
total pressure and is thus related to the plasma physics $\beta$ used
in this paper by $\beta_{\rm adv} = \beta/(\beta + 1)$.}, $\op$ is the
proton cyclotron frequency, $\rp$ is the Larmor radius of thermal
protons, $m$ is the mass of the central object in solar mass units, $M
= m \ M_{\odot}$, $\dot{m} $ is the accretion rate in Eddington units,
$\dot{M} = \dot{m} \ \dot{M}_{\rm Edd}$ ($\dot{M}_{\rm Edd} =
1.39\times 10^{18} m$ g s$^{-1}$), and $r = R/R_S$ is the radius in
Schwarzschild units ($R_S = 2.95 \times 10^5 m$ cm).  In equation
(\ref{adaf}), the fraction of the viscously dissipated energy that is
carried inward by the accreting gas is taken to be $\sim 1$.

The electron temperature in ADAF models typically saturates at $T_e
\sim 10^9-10^{10}$ K in the inner $10^2-10^3$ Schwarzschild radii
since the efficient cooling of relativistic electrons prevents higher
temperatures.  Thus the electrons and protons are both marginally
relativistic and the non-relativistic analysis employed in this paper
is a reasonable first approximation.  Using equation (\ref{adaf}) we
can compute the characteristic frequency ($\nu$) and (field-free) mean
free path ($\ell$) for proton-proton ($pp$) and electron-electron
($ee$) Coulomb collisions (Mahadevan \& Quataert 1997): \beq \nu_{pp}
\simeq 30 \alpha^{-1} \left( {\beta +1 \over \beta }\right)^{3/2}
m^{-1} \dot m \ \ \mbox{Hz}, \eeq \beq \nu_{ee} \simeq 10^8
\alpha^{-1} m^{-1} \dot m r^{-3/2} T^{-3/2}_9 \ \ \mbox{Hz},
\label{nuee} \eeq 
\beq {\ell_{pp} \over R} \simeq 10^3 \alpha \left({\beta \over \beta +
1}\right)^2 {\dot m}^{-1} r^{-3/2}, \eeq \beq {\ell_{ee} \over R}
\simeq 10^{-3} \alpha {\dot m}^{-1} r^{1/2} T^2_9, \eeq where $T_9$ is
the electron temperature in units of $10^9$ K.  $\nu_{ee}$
($\nu_{pp}$) is the rate at which the energy and direction of an
electron (proton) changes appreciably through Coulomb collisions with
the same particle species.  Electron-proton collisions change the
proton energy at a rate $ \simeq \nu_{ee} m_e/m_p $.  $\ell$ is the
mean free path in the absence of a magnetic field.  Since $\ell$ is
often $\gsim R$, this is indicative of the collisionless nature of the
plasma in ADAFs.

The characteristic frequency at the outer scale of the accretion flow,
which is roughly the frequency at which waves (both Alfven and slow
and fast MHD modes) will be excited, is $\om_{out} \simeq v_A R^{-1}
\simeq 6 \times 10^{4} (\beta + 1)^{-1/2} m^{-1} r^{-3/2}$ rad
s$^{-1}$. Comparing this with $\nu_{pp}$, we see that the protons are
effectively collisionless for all perturbations of interest (those
with frequencies $\gsim \om_{out}$).  By contrast, however, \beq
\nu_{ee}/\om_{out} \sim 10^3 \alpha^{-1} (\beta + 1)^{1/2} \dot m
T_9^{-3/2}.
\label{out} \eeq
Provided that \beq \dot m \gsim 10^{-3} \alpha (\beta + 1)^{-1/2}
T^{3/2}_9, \label{md} \eeq $\nu_{ee} \gsim \om_{out}$ and the
electrons must be treated as {\em collisional} on the outer
scale. This is probably not a significant complication for Alfven
waves, since collisionless effects are unimportant at the outer
scale. It must, however, be taken into account in treatments of the
fast and slow modes, which undergo collisionless dissipation even in
the MHD limit (\S3.5).  The fast mode is damped primarily by the
protons.  The net damping of the wave will therefore not be
significantly modified by the collisionality of the electrons.  The
slow mode, on the other hand, is a modified sound wave which is
strongly Landau damped by electrons in a collisionless plasma with
$T_p \gg T_e$.  The collisionality of the electrons for perturbations
with $\om \sim \om_{out}$ will suppress the electron contribution to
the damping and thus remove this source of electron heating (for
accretion rates satisfying eq. [\ref{md}]).  The slow mode will,
however, still be strongly damped by other mechanisms at the outer
scale, e.g., Landau damping by the protons and wave steepening leading
to (collisionless) shocks.

For the GS cascade, strong damping of the Alfven wave occurs when $\kp
\simeq \rp^{-1}$, at which point $k_z \simeq \rp^{-2/3} R^{-1/3}$ and
$\om \equiv \om_{in} \simeq \op \beta^{-1/2} (\rp/R)^{1/3}$.  Using
equation (\ref{adaf}) and comparing $\om_{in}$ with $\nu_{ee}$, we
find that \beq \om_{in}/\nu_{ee} \simeq 10^{2} \alpha^{2/3}
\beta^{-1/2} (\beta + 1)^{-1/2} m^{1/3} {\dot m}^{-2/3} r^{1/6}
T^{3/2}_9. \eeq Thus, in the dissipation range, collisionless theory
can be consistently applied for both electrons and protons.  This
conclusion is strengthened by noting that $m \gsim 1$ and $\dot m \lsim
1$.

\subsection{Collisional Dissipation of Alfven Waves}

The (field free) mean free path for Coulomb collisions in a hot
accretion flow is a significant fraction of the local radius.  This
might suggest that dissipation of waves by microscopic viscosity could
be important.  The velocity fluctuation of an Alfven wave is, however,
perpendicular to the local magnetic field and so only cross field
components of the viscosity tensor are important (GS).  These are
smaller than the field free component by a factor of $\sim
(\nu_{pp}/\op)^2 \ll 1$, making microscopic viscous dissipation
unimportant.  More concretely, the dissipation range of a turbulent
cascade set by microscopic viscous effects occurs at a scale $r_v \sim
R/Re^{3/4}$, where the Reynold's number is $Re \sim v_A R
/\mu_{\perp}$ and $\mu_{\perp} \simeq 0.1 n/(T^{1/2}_p B^2)$ cm$^{-2}$
s$^{-1}$ is the cross field kinematic viscosity coefficient (e.g.,
Spitzer 1961).  Using equation (\ref{adaf}) we find that $Re \sim
10^{21} \beta^{1/2} (\beta + 1)^{-2} m r^{-5/4}$ and so \beq r_v/\rp
\sim 10^{-7} \alpha^{-1/2} \beta^{-7/8} (\beta + 1)^{3/2} m^{-1/4}
{\dot m}^{1/2} r.  \eeq Microscopic viscous effects thus become
important only on negligibly small scales.  An analogous result holds
for thermal conductivity.

Finite electrical resistivity also leads to wave damping, which
becomes important at a scale $r_{\eta_e} \sim R/(Re_m)^{3/4}$, where
$Re_m \sim v_A R/\eta_e$ is the Magnetic Reynold's number at the outer
scale and $\eta_e \simeq 10^{13}/T^{3/2}_e \sim T_9^{-3/2}$ cm$^2$
s$^{-1}$ is the electrical resistivity (Spitzer 1961).  Using equation
(\ref{adaf}) yields $Re_m \sim 10^{16} \beta^{-1/2} m r^{-1/2}
T^{3/2}_9$ and so resistive damping occurs when \beq r_{\eta_e}/\rp
\sim 10^{-4} \beta^{-1/8} m^{-1/4} {\dot m}^{1/2} r^{1/8}
T_9^{-9/8}. \eeq Collisionless dissipation sets in on scales $\sim
\rp$, which is $\gg$ than the collisional dissipation scales
considered in this section; as the energy cascades to small
wavelengths, it will therefore first encounter collisionless
dissipation processes.

\subsection{Implications of Particle Heating for ADAFs}

It is usual in ADAF models to specify the relative heating of protons
and electrons by a parameter $\delta$, the fraction of the viscous
energy which heats the electrons.  In order for the optically thin
ADAF formalism to be relevant to an accretion flow, one of two
physical situations must occur: 

{\noindent 1.  $\delta \lsim 0.5$ and ${\dot m} \lsim \alpha^2$.  In
this case a significant fraction of the viscous energy is transferred
to the protons; by virtue of the low accretion rate (which implies low
densities; see eq. [\ref{adaf}]), Coulomb collisions are too
inefficient to transfer this energy to the electrons in the inflow
time of the gas (Rees et al. 1982).  Consequently, a fraction $\sim 1
- \delta$ of the viscous energy is advected, by the protons, onto the
central object.}

{\noindent 2.  $\delta \sim 1$ and $\dot m \lsim 10^{-4} \alpha^2$.
In this case most of the viscous energy heats the electrons; the flow
can be advection dominated only if the electron cooling time (the time
for the electrons to radiate their thermal energy) is longer than the
inflow time of the gas.  For synchrotron cooling, the most efficient
cooling mechanism in ADAFs at low accretion rates (Narayan \& Yi
1995b), this occurs for ${\dot m} \lsim 10^{-4} \alpha^2$ (Mahadevan
\& Quataert 1997).

The contribution to $\delta$ from the collisionless dissipation of
Alfven waves considered in \S3 is (eq. [\ref{aheat}]) \beq \delta
\equiv {P_e \over P_p + P_e} \sim \left(m_e T_e \over m_p T_p
\right)^{1/2} \exp\big[(1 + T_e/T_p) \beta^{-1} \big].
\label{delta} \eeq For accretion flows with $T_p \gg
T_e$, this contribution is $\ll 1$.  For most ($\gsim 50 \%$) of the
energy in Alfvenic turbulence to be dissipated by the mechanisms
considered in this paper, however, the accretion flow must have $\beta
\gsim 5$ (\S5.2), which is different from the value of $\beta \sim 1$
usually used in ADAF models.  Furthermore, numerical simulations of
MHD turbulence in thin disks suggest that $\alpha$ and $\beta$ are
coupled, with $\alpha \simeq 0.5/(\beta + 1)$ (Hawley, Gammie, \&
Balbus 1996; Table 4).  In this case $\beta \gsim 5$ would correspond
to $\alpha \lsim 0.1$, which is smaller than the value of $\alpha
\simeq 0.25$ usually used in ADAF models.  This $\alpha-\beta$
relationship may not, however, be applicable to ADAFs, which are {\em
radially} convective and thus have a purely hydrodynamic source of
angular momentum transport.

ADAF models typically take $\delta \sim m_e/m_p \sim 10^{-3}$ (e.g.,
Narayan et al. 1997), but they are relatively insensitive to $\delta$
so long as $\delta \ll 1$.  The reason is that, in this limit, viscous
heating of electrons is not their dominant heating mechanism.  This
can be seen by considering the energy equation for electrons in a hot
accretion flow, \beq \rho T v {ds \over dR} = \rho v {d \epsilon \over
dR} - q^c = q^{\rm e+} - q^-,
\label{eeq}
\eeq
$$ q^c \equiv kT v {dn \over dR}, $$
$$q^{\rm e+} = q^{\rm ie} + q^v, $$ where $s$ is the entropy of the
electrons per unit mass of the gas, $\epsilon$ is the internal energy
of the electrons per unit mass, $q^c$ is the compressive heating (or
cooling) rate per unit volume, and $q^{-}$ is the energy loss due to
radiative cooling.  The total external heating of the electrons,
$q^{\rm e+}$, is a sum of the heating via Coulomb collisions with the
hotter protons, $q^{\rm ie}$, and direct viscous heating, $q^v$.

For accretion rates such that \beq \dot m \gsim 10^{-4} \alpha^2 \left
(\delta \over 10^{-3} \right) T^{3/2}_9,
\label{mdot}\eeq  Coulomb heating of the electrons dominates over viscous 
heating (Mahadevan 1997), and so the precise value of $\delta$ is
unimportant.  The ratio of compressive to viscous heating is given
roughly by (Mahadevan \& Quataert 1997) \beq q^c/q^v \sim 10^{-3}
\delta^{-1} T_9 r. \label{comp}\eeq For small $\delta$, compressive
heating of the electrons is more important than viscous heating, again
making the precise value of $\delta$ unimportant.  For most of the
systems to which ADAF models have been applied, $\dot m$ is
sufficiently high and compressive heating is sufficiently important
that only if $\delta$ were $\gsim 0.03$ in the inner regions of the
accretion flow ($r \sim 3-100$, where the observed radiation
originates) would the models be significantly modified.\footnote{An
important exception to this is the Narayan et al. (1997) model of
Sagittarius A$^*$.  For this system, the data is so good and the
estimated accretion rate is so low ($\dot m \sim 10^{-4}$) that
increasing $\delta$ to $\gsim 10^{-2}$ is problematic (see Fig. 4 of
their paper).}

\subsection{Isotropy?}

The mechanisms for dissipating Alfvenic turbulence considered in this
paper heat only the component of a particle's energy which is parallel
to the local magnetic field (\S4).  In our calculations, however, we
have assumed an isotropic distribution function, which must be
justified.  A standard isotropization mechanism for both electrons and
protons is pitch angle scattering. This can occur much more
efficiently than particle heating if the parallel phase speed of the
waves responsible for the scattering is much less than the particle's
speed (Melrose 1980); in this case isotropy could be maintained
without significant heating.  While this is a plausible mechanism for
maintaining isotropy in hot accretion flows, we have not investigated
this mechanism in detail.

Rather, we wish to point out a mechanism, unique to ADAFs, which
maintains rough isotropy in the proton distribution function provided
that the turbulent heating mechanism affects only the parallel
component of the proton's energy.  In addition to heating by the
viscously generated energy, particles are heated by compression as
they accrete inwards.  Since the particles are tied to the field
lines, which are being compressed, adiabatic invariance of the
particle's magnetic moment requires $E_{\perp} \propto B$, where
$E_{\perp}$ is the particle's perpendicular energy.  For protons in an
ADAF, the viscous heating rate is comparable to the compressional
heating rate; this is because most of the viscous energy is stored as
thermal energy of the protons and so all terms in the proton entropy
equation are of the same order.  If the viscous heating mechanism
heats primarily the parallel energy of the protons, as the
wave-particle interactions considered in this paper do, then the rough
equality of viscous and compressional heating rates implies a rough
equality of parallel and perpendicular heating, so that isotropy is
maintained.  Furthermore, Mahadevan \& Quataert (1997) have shown that
adiabatic compression maintains a thermal distribution of particles
even in a collisionless gas (provided the particles are not compressed
to relativistic energies).  Therefore, given a thermal distribution of
protons at large radii in an accretion flow, the perpendicular
component of the distribution function remains thermal as they accrete
onto the central object.

This mechanism will not work for electrons since there is no necessary
relationship between the compressive and viscous heating rates (the
electrons are not, in general, advection dominated).  For accretion
rates satisfying equation (\ref{md}), however, Coulomb collisions can
maintain isotropy in the electron distribution function.  This is
because the characteristic timescale on which Alfvenic turbulence
modifies the proton and electron distribution functions is $\sim
\om_{out}^{-1}$.\footnote{This can be estimated from the quasi-linear
diffusion equations, using the results of \S5.1 (i.e., it is only the
parallel fields, which arise from kinetic corrections to GS's MHD
cascade, which lead to diffusion of the particles in velocity space).}
Since the electrons (but not the protons) are effectively collisional
on this timescale, isotropy is maintained.

\subsection{Comments on related work}

In a recent paper, Bisnovatyi-Kogan \& Lovelace (1997; hereafter BL)
suggest a mechanism which they feel leads to preferential electron
heating in hot accretion flows.  It is worthwhile examining this in
some detail.

Throughout this paper we have argued that, on a microscopic level,
particle heating occurs when the MHD turbulence generated at large
scales in the accretion flow cascades to small scales and is
dissipated by collisionless effects.  BL seem to disregard this
possibility, instead claiming that because the dissipation scale set
by the hydrodynamic or magnetic Reynold's number is extremely small,
plasma instabilities set in which dissipate the turbulent energy.
Collisionless effects, however, become important on length scales well
above those set by the hydrodynamic or magnetic Reynold's number (\S
6.2), obviating the need for an appeal to plasma instabilities.

BL argue that, on a microscopic level, particle heating in ADAFs is
due to electric fields parallel to the local magnetic field
accelerating particles to runaway velocities.  By virtue of their
smaller mass, electrons are more efficiently accelerated.  In MHD,
however, the electric field is given by ${\bf E = - v \times B}/c +
\eta_e {\bf \nabla \times B}/4 \pi c$.  In a highly conducting plasma,
the magnitude of a typical electric field is $\sim v B/c$.
Furthermore, the electric field is primarily perpendicular to the
local magnetic field and thus unimportant for accelerating particles.
In MHD parallel electric fields arise only from finite resistivity
corrections and are $\sim v B/{ c Re_m}$; since $Re_m \gg 1$ they are
far too small to significantly accelerate particles in hot accretion
flows.

Following Bisnovatyi-Kogan \& Ruzmaikin (1976), however, BL argue
that, because the flow is turbulent, one should use a ``turbulent
resistivity'' (of order the Shakura-Sunyaev turbulent viscosity)
instead of the usual microscopic resistivity; in this case, they
argue, $\Ep \sim v B/c$ and parallel electric fields can significantly
accelerate particles.  Our concern with this analysis is that, while
the notion of a turbulent transport coefficient may be useful in
describing the large scale properties of the flow (the global
transport of angular momentum, for example), it should not be used on
a microscopic level.  When considering microscopic processes such as
particle acceleration, the global transport processes in the flow are
unimportant.

BL's electron heating mechanism amounts to the claim that, on a
microscopic level, the local electric field is {\em significantly
aligned} with the local magnetic field.  We see no reason, however,
why the MHD results described above (the local electric field {\em
perpendicular} to the local magnetic field) should be inapplicable to
hot accretion flows.

\section{Summary and Discussion}

Hot accretion flows, such as advection dominated accretion flows
(ADAFs), are effectively collisionless for all but the largest scale,
lowest frequency, motions of the plasma (\S6.1).  Collective plasma
effects are thus likely to be important in these systems.  This is
particularly true for particle heating since microscopic viscosity,
thermal conductivity, and electrical resistivity are important only at
extremely small scales.  Particle heating influences both the global
structure of the flow (e.g., by determining if the protons or
electrons are heated) as well as the observed radiation (e.g., by
determining the particle distribution functions).

Particle heating is a particularly important issue for ADAF
models. All ADAF models which have been applied to observed systems
assume that the viscously generated energy primarily heats the
protons.  This enables the accretion flow to be advection dominated so
long as the timescale for electrons and protons to exchange energy by
Coulomb collisions is longer than the inflow time of the gas.  If
viscosity {\em only} heats the electrons, an (optically thin)
accretion flow can be advection dominated only when the electron
cooling time is longer than the inflow time, which occurs at such low
accretion rates that ADAF models would probably be less relevant for
observed systems.  Recent work by Bisnovatyi-Kogan \& Lovelace (1997)
suggests that all of the viscous energy in hot accretion flows does
heat the electrons.  In \S6.5 we have argued against this conclusion.

In this paper we have focused on one aspect of the particle heating
problem, namely particle heating by the linear collisionless
dissipation of MHD, in particular Alfven, waves.  We find it likely
that the viscous energy in hot accretion flows resides primarily in
MHD turbulence, making this mechanism of particular importance.  An
investigation similar to ours, but with somewhat different
conclusions, was carried out independently by Gruzinov (1997).

Long wavelength Alfven waves excited on the outer scale in hot
accretion flows are undamped by linear collisionless effects.  The
nature of Alfvenic turbulence is thus crucial for assessing Alfven
wave damping and particle heating.  Recent work on Alfvenic turbulence
by Goldreich and Sridhar (1995; GS) suggests that the turbulent energy
cascades almost entirely perpendicular to the local magnetic field
(i.e., in the inertial range, $\kp \gg k_z$).  As a result, the
cyclotron resonance, which is usually thought to be significant in
dissipating Alfvenic turbulence, is unimportant (since the wave
frequencies are always much less than the proton cyclotron frequency).

We have shown, using the full kinetic theory dispersion relation for
linear perturbations to a plasma, that Alfven waves with frequencies
much less than the proton cyclotron frequency, but with perpendicular
wavelengths of order the Larmor radius of thermal protons, are damped
by transit time damping (TTD) in plasmas appropriate to hot accretion
flows ($T_p \gsim T_e$ and $\beta$, the ratio of the gas pressure to
the magnetic pressure, $\gsim 1$).  TTD is due to particles being
accelerated by gradients in a wave's longitudinal magnetic field
perturbation, and is the magnetic analogue of Landau damping. For $T_p
\gg T_e$, TTD quite generally leads to most of the dissipated wave
energy heating the protons, since they have the larger magnetic moment
and so couple better to the wave's magnetic field perturbation (\S3.3
and \S3.4).  For example, if $T_p \simeq 100 T_e$, as is the case in
the interior of ADAF models, the electrons receive only $\sim 2 \%$ of the
Alfven wave energy as it is damped in a $\beta \sim 1$ plasma.

The dissipation rate of the Alfven wave increases with increasing
$\beta$ since the proton's magnetic moment is effectively larger and
the wave-particle coupling is stronger (Figure 2; \S3.3).  For $\beta
\gsim 5$, the dissipation of Alfven waves by TTD is sufficiently
strong to convert most ($\gsim 50 \%$) of the Alfvenic energy to
thermal energy before the MHD approximations utilized in the GS
analysis cease to be valid.  In the GS cascade, however, the energy
travels so quickly through the inertial range that, for $\beta \sim
1$, TTD is not strong enough to dissipate the turbulent energy.  In
this case, the GS analysis of the turbulent dynamics breaks down
before most of the turbulent energy is dissipated, and it is unclear
to us how the cascade proceeds; how the Alfvenic energy is ultimately
dissipated in a $\beta \sim 1$ plasma therefore remains unresolved by
our work. In $\S5.2$ we have enumerated (qualitatively) what we take
to be the plausible possibilities.

So long as a significant fraction ($\gsim 50 \%$) of the viscous
energy heats the protons, the ADAF formalism is capable of describing
the structure of an accretion flow (\S6.3). This work suggests that,
for $\beta \gsim 5$, Alfvenic turbulence in hot, two temperature,
accretion flows preferentially heats the protons.  Since we expect a
significant fraction of the viscously generated energy to reside in
Alfvenic turbulence, this alone can plausibly lead to greater than
$\sim 50$ \% of the total viscously generated energy heating the
protons.

Wave-particle interactions are often suggested as a mechanism for
forming strong nonthermal features in the electron and proton
distribution functions.  In a perpendicular Alfvenic cascade of the
kind proposed by GS, however, the wave damping is always due to
particles with velocities equal to the Alfven speed (\S4).  In linear
theory, it is thus impossible to accelerate particles to
supra-Alfvenic velocities.  For $\beta \sim 1$ and $T_p \gg T_e$, the
Alfven speed is near the thermal peak of the proton and electron
distribution functions, so that the dissipated turbulent energy does
not significantly modify the distribution functions from a Maxwellian.
Perpendicular Alfvenic turbulence is therefore {\em not} a plausible
mechanism for producing power law features in the proton and electron
distribution functions or for accelerating particles to relativistic
energies. This may exclude Alfvenic turbulence as a viable mechanism
for particle acceleration in solar flares and accretion disk corona.

\noindent{\it Acknowledgments.} I would like to thank Ramesh Narayan
for numerous useful conversations and Andrei Gruzinov for sending me a
draft of his paper prior to publication.  Anthony Aguirre, Bruno
Coppi, Ann Esin, George Field, Charles Gammie, Zoltan Haiman, Rohan
Mahadevan, and Kristen Menou also provided helpful comments on this
work.  EQ was supported by an NSF Graduate Research Fellowship.  This
work was also supported in part by NSF Grant AST 9423209.

\newpage

\begin{appendix}

\section{Definitions of Some Oft-used Quantities}
\begin{tabular}{ccc} 
Quantity  & Definition & Meaning\\
\hline
$\omega$ & & mode frequency \\
$T$ & 2 $\pi$/Re$(\omega)$ & mode period \\
$\gamma$ & Im$(\omega)$ & mode damping rate \\ 
{\bf k} & & mode wavevector \\
$\Omega_p$ & $q B/m_p c$ & proton cyclotron frequency \\
$v_{ts}$ & $\sqrt{2k_B T_s/m_s}$ & proton (s = p) and electron (s = e) thermal speeds \\
$\rp$ & $v_{tp}/\Omega_p$ & Larmor radius of thermal protons \\
$\lp$ & $0.5 \kp^2 \rp^2$ & dimensionless perpendicular wavevector \\
$\eta$ & $k_z \rp$ & dimensionless parallel wavevector \\
$\alpha$ & & Shakura-Sunyaev viscosity parameter \\
$\beta$ & $8 \pi n k_B(T_p + T_e)/B^2$ & ratio of gas to magnetic pressure \\
$P_s$ & & proton (s = p) and electron (s = e) heating rates \\
$\delta$ & & fraction of viscous energy heating electrons \\
\hline
\end{tabular}

\section{The Balbus-Hawley instability in a collisionless gas}
There are three points which suggest to us that the MHD instability of
Balbus \& Hawley (BH) should apply to collisionless systems.  While
suggestive, these do not, of course, constitute a proof.  The primary
caveat to these comments is that, if the particle distribution
functions are highly nonthermal, all bets are off.

{\noindent 1.  Perhaps the primary concern in passing from the
collisional to the collisionless version of an instability is that
collisionless dissipation mechanisms may inhibit the instability.  To
linear order the axisymmetric version of the BH instability is,
however, non-compressive and Alfvenic in character.  As discussed in
\S3.1, in the MHD limit Alfven waves are undamped by linear
collisionless effects, which suggests that the instability should not
be inhibited.}

{\noindent 2.  The collisionless limit entails the infinite
conductivity limit used in ideal MHD; finite resistivity effects are
particularly unimportant in a collisionless plasma.}

{\noindent 3. The Keplerian rotation frequency, which is the
characteristic growth rate of the instability, is $\Omega_o \simeq 7
\times 10^4 m^{-1} r^{-3/2}$ rad s$^{-1}$.  The smallest
characteristic frequency in a plasma is typically the proton cyclotron
frequency, which is given in equation (\ref{adaf}).  For $\beta \ll
10^{16}$, $\Omega_o \ll \Omega_p$ and so the particles are tied to the
field lines, which is a requirement for the instability to function.
In the limit of $\Omega_o \gsim \op$ it is unlikely that the
instability will persist.  This corresponds, however, to exceedingly
small magnetic field strengths ($\beta \gsim 10^{16}$).}

\end{appendix}

\newpage

{
\footnotesize
\StartRef
\noindent {\large \bf References} \\
\Ref Abramowicz, M., Chen, X., Kato, S., Lasota, J. P, \& Regev, O., 1995,
ApJ, 438, L37 \\
\Ref Achterberg, A. 1981, A\&A, 97, 259 \\
\Ref Akhiezer, A. I., Akhiezer, I. A., Polovin, R. V., Sitenko, A.G., \& Stepanov, K.N. 1975, Plasma Electrodynamics Volume I  (Pergamon, Oxford, 1975) \\
\Ref Balbus, S. A., \& Hawley, J.F. 1991, ApJ, 376, 214 \\
\Ref Balbus, S. A., \& Hawley, J.F. 1997, Rev. Mod. Phys., in press \\
\Ref Barnes, A. 1966, Phys. Fluids, 9, 1483 \\
\Ref Barnes, A. 1967, Phys. Fluids, 10, 2427 \\
\Ref Barnes, A. 1968a, ApJ, 154, 751 \\
\Ref Barnes, A. 1968b, Phys. Fluids, 11, 2644 \\
\Ref Begelman, M. C. \& Chiueh, T. 1988, ApJ, 332, 872 \\
\Ref Bisnovatyi-Kogan, G.S. \& Lovelace, R. V. E. 1997, ApJ, 486, L43 \\
\Ref Bisnovatyi-Kogan, G.S. \& Ruzmaikin, A. A. 1976, Ap\&SS, 42, 401 \\
\Ref Chen, X., Abramowicz, M.A., Lasota, J.-P., Narayan, R., \& Yi, I. 1995, ApJ, 443, L61 \\
\Ref Dermer, C. D., Miller, J. A., \& Li, H. 1996, ApJ, 456, 106 \\
\Ref Frank, J., King, A., \& Raine, D., 1992, Accretion Power in
Astrophysics (Cambridge: Cambridge Univ. Press) \\
\Ref Foote, E. A. \& Kulsrud, R. M. 1979, ApJ, 233, 302 \\
\Ref Goldreich, P. \& Sridhar, S. 1995, ApJ, 438, 763 \\
\Ref Goldreich, P. \& Sridhar, S. 1997, ApJ, 485, 680 \\
\Ref Gruzinov, A. 1997, ApJ submitted \\
\Ref Hasegawa, A. \& Chen, L. 1976, Phys. Fluids, 19, 1924 \\
\Ref Hawley, J. F., Gammie, C. F., \& Balbus, S. A. 1996, ApJ, 464, 690 \\
\Ref Hollweg, J. V. 1971 PRL, 27, 1349 \\
\Ref Ichimaru, S. 1977, ApJ, 214, 840 \\
\Ref Kato, S., Abramowicz, M., \& Chen, X. 1996, PASJ, 48, 67 \\
\Ref Kraichnan, R.H. 1965, Phys. Fluids, 8, 1385 \\

\Ref Li, H., Kusunose, M., \& Liang, E. 1996, ApJ, 460, L29 \\
\Ref Li, H. \& Miller, J. A. 1997, ApJ, 478, L67 \\
\Ref Mahadevan, R., 1997, ApJ, 477, 585-601 \\ 
\Ref Mahadevan, R. \& Quataert, E. 1997, ApJ, 490, in press \\
\Ref Melrose, D.B. 1980, Plasma Astrophysics (New York: Gordon \& Breach) \\
\Ref Melrose, D.B. 1986, Instabilities in space and laboratory plasmas (New York: Cambridge University Press) \\
\Ref Melrose, D.B. 1994, ApJS, 90, 623 \\
\Ref Miller, J. A. 1991, ApJ, 376, 342 \\
\Ref Miller, J. A. \& Roberts, D. A. 1995, ApJ, 452, 912 \\
\Ref Miller, J. A., LaRosa, T.N., \& Moore, R. L. 1996, ApJ, 461, 445 \\
\Ref Montgomery, D. \& Matthaeus, W. H. 1995, ApJ, 447, 706 \\ 
\Ref Nakamura, K. E., Masaaki, K., Matsumoto, R., \& Kato, S. 1997, PASJ, in press \\
\Ref Narayan, R., \& Yi, I., 1994, ApJ, 428, L13 \\ 
\Ref Narayan, R., \& Yi, I., 1995a, ApJ, 444, 231 \\ 
\Ref Narayan, R., \& Yi, I., 1995b, ApJ, 452, 710 \\ 
\Ref Narayan, R., Mahadevan, R., Grindlay, J.E., Popham, R.G., \& Gammie, C., 1997, ApJ, in press \\
\Ref Ng, C. S. \& Bhattacharjee, A. 1996, ApJ, 465, 845 \\ 
\Ref Phinney, E. S. 1981, in Plasma Astrophysics, ed. T Guyenne (ESA SP-161), p. 337 \\
\Ref Rees, M. J., Begelman, M. C., Blandford, R. D., \& Phinney,
E. S., 1982, Nature, 295, 17 \\

\Ref Shakura, N. I., \& Sunyaev, R. A., 1973, A\&A, 24, 337 \\
\Ref Shapiro, S. L., Lightman, A. P., \& Eardley, D. M. 1976, ApJ, 204,
 187 \\ 
\Ref Shebalin, J. V., Matthaeus, W. H., \& Montgomery, D. 1983, J. Plasma Phys., 29, 525 \\
\Ref Smith, D. F. \& Miller, J. A. 1995, ApJ, 446, 390 \\
\Ref Spitzer, L. Jr., 1962, Physics of Fully Ionized Gases,
2nd Ed., (New York: John Wiley \& Sons, Inc.) \\ 
\Ref Sridhar, S. \& Goldreich, P, 1994, ApJ, 432, 612 \\
\Ref Stefant, R. 1970, Phys. Fluids, 1, 440 \\
\Ref Stone, J. M., Hawley, J. F., Gammie, C. F., Balbus, S. A. 1996, ApJ, 463, 656 \\
\Ref Stix, T.H. 1992, Waves in Plasmas (New York: AIP) \\ 
}
\newpage
\noindent{\bf Figure Captions.} \\

\noindent Figure 1: Properties of the Alfven wave as a function of
$\lp = 0.5 \kp^2 \rp^2$ for $\om \ll \op$; various proton to electron
temperature ratios, $T_p/T_e$, are considered for a plasma with equal
gas and magnetic pressure ($\beta = 1$).  (a) The parallel phase speed
in units of the Alfven speed.  For $T_p \gsim 10 T_e$, $\vp$ is nearly
the same as for $T_p = 10^3 T_e$.  (b) The dissipation per mode
period, $\gamma T$.  (c) The dimensionless proton (solid line) and
electron (dotted line) heating rates ($P_s$).  $T_p/T_e$ is shown
along side each curve.  For $T_p \gsim 10 T_e$, the proton heating
rate, $P_p$, is nearly identical to the $T_p = 10^3 T_e$ case. (d) The
relative proton and electron heating rates, $P_p/P_e$.

\noindent Figure 2: Properties of the Alfven wave as a function of
$\lp = 0.5 \kp^2 \rp^2$ for $\om \ll \op$; various ratios of gas
pressure to magnetic pressure ($\beta$) are considered for a $T_p =
100T_e$ plasma.  (a) The dissipation per mode period, $\gamma T$.  (b)
The relative proton and electron heating rates, $P_p/P_e$.

\noindent Figure 3: The relative contribution of transit time damping
($\chi^a_{yy} |E_y|^2$) and Landau damping ($\chi^a_{zz} |E_z|^2$) to
the (a) proton and (b) electron heating rates as a function of the
proton to electron temperature ratio ($T_p/T_e$); several $\beta$ are
considered, taking $\lp = 0.1$.

\noindent Figure 4: The fractional dissipation per mode period,
$\gamma T$, in the MHD limit for (a) the fast mode and (b) the slow
mode as a function of the angle between the wavevector and the
background magnetic field ($\theta$); various $T_p/T_e$ are considered
for a $\beta = 1$ plasma.  For the slow mode, the curves are nearly
vertical displacements of each other, so fewer $T_p/T_e$ are shown.

\newpage
\begin{figure}
\epsffile{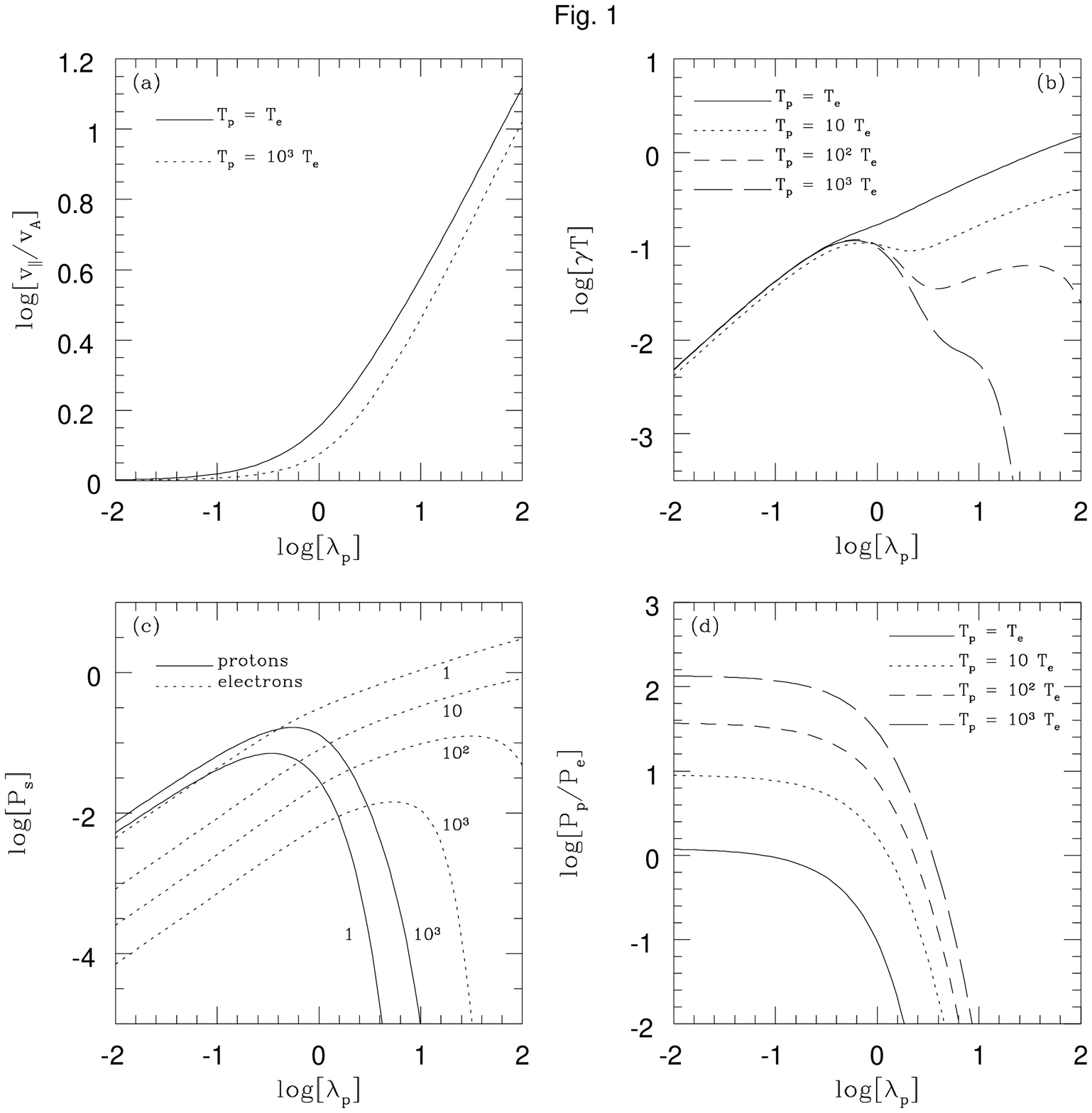}
\end{figure}
\begin{figure}
\epsffile{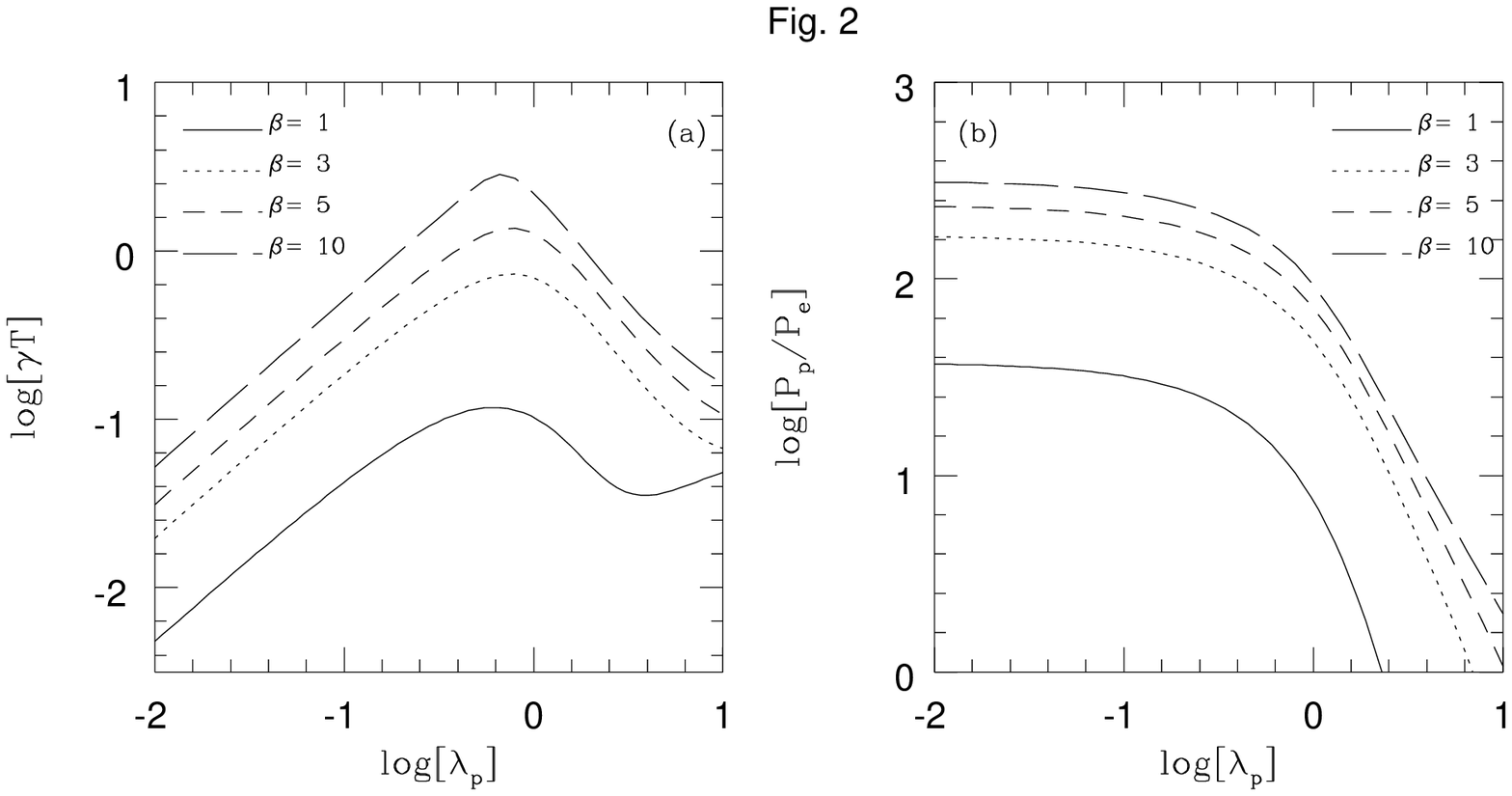}
\end{figure}
\begin{figure}
\epsffile{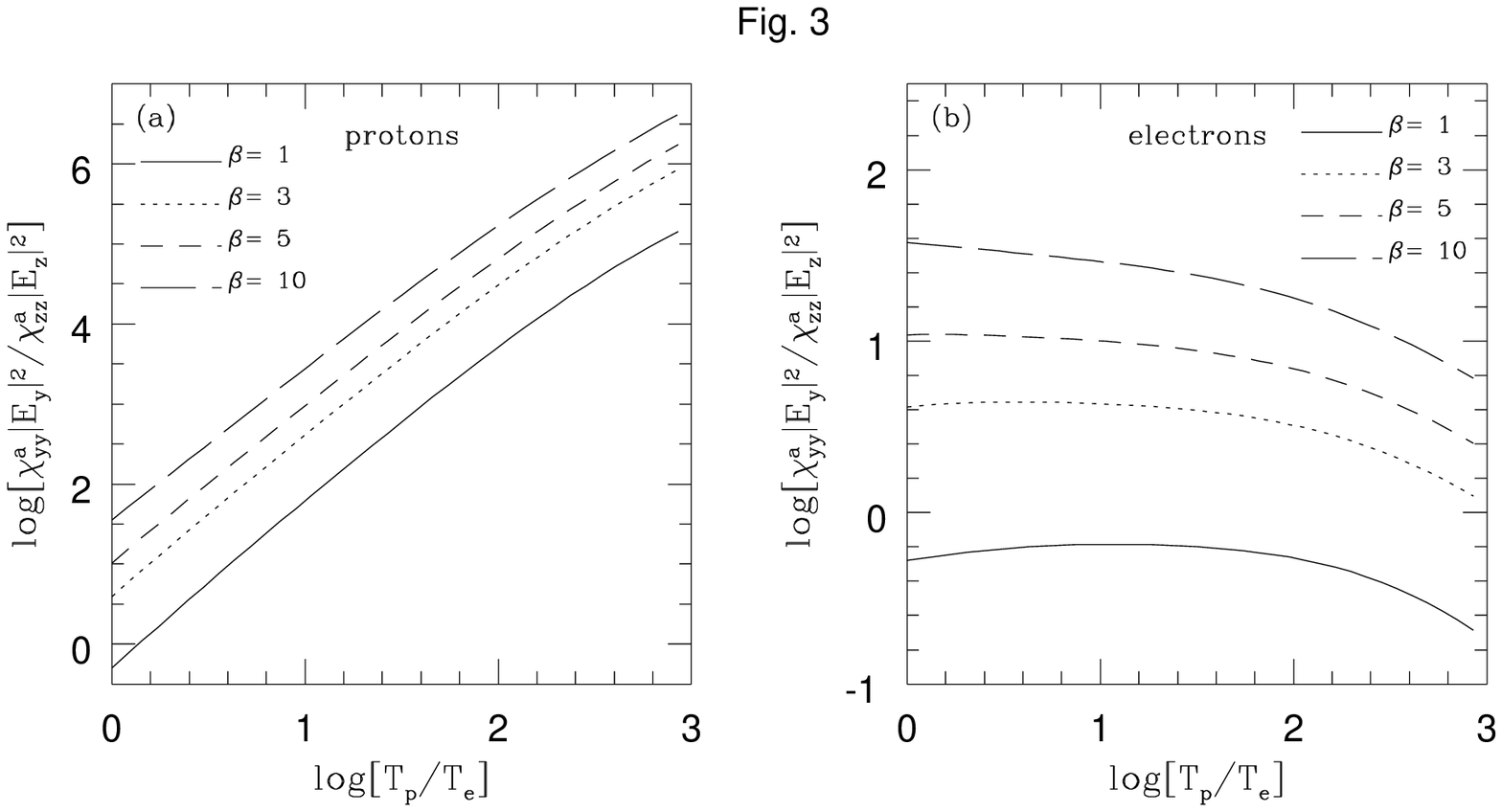}
\end{figure}
\begin{figure}
\epsffile{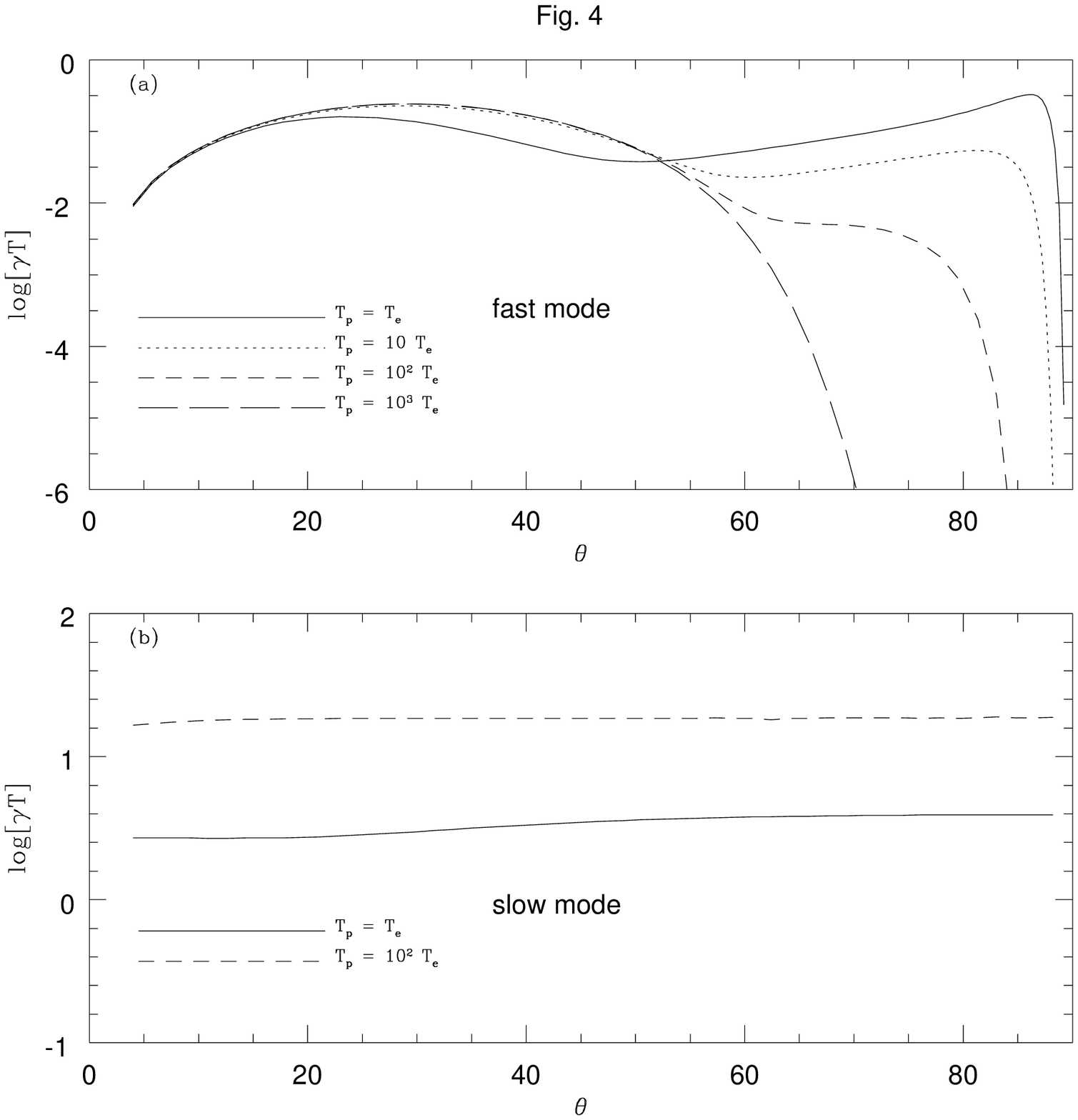}
\end{figure}

\end{document}